# Ultra-Fast Fluorescence Imaging *in Vivo* with Conjugated Polymer Fluorophores in the Second Near-Infrared Window

Guosong Hong[1,6], Yingping Zou[2,4,5,6], Alexander L. Antaris[1,6], Shuo Diao[1], Di Wu[1], Kai Cheng[3], Xiaodong Zhang[1], Changxin Chen[1], Bo Liu[2,4], Yuehui He[4], Justin Z. Wu[1], Jun Yuan[2], Bo Zhang[1], Zhimin Tao[1], Chihiro Fukunaga[1], and Hongjie Dai[1]

[1] Department of Chemistry, Stanford University, Stanford, California 94305, USA

[2] College of Chemistry and Chemical Engineering, Central South University, Changsha 410083, China

[3] Department of Radiology and Bio-X Program, School of Medicine, Stanford University, Stanford, California 94305, USA

[4] State key Laboratory for Powder Metallurgy, Central South University, Changsha 410083, China

[5] Institute of Super-microstructure and Ultrafast Process in Advanced Materials, School of Physics and Electronics, Central South University, Changsha 410083, China

[6] These authors contributed equally to this work.

Correspondence should be addressed to: hdai@stanford.edu and yingpingzou@csu.edu.cn

## Abstract

*In vivo* fluorescence imaging in the second near-infrared window (1.0-1.7 μm) can afford deep tissue penetration and high spatial resolution, owing to the reduced scattering of long-wavelength photons. Here, we synthesize a series of low-bandgap donor/acceptor copolymers with tunable emission wavelengths of 1050-1350 nm in this window. Non-covalent functionalization with phospholipid-polyethylene glycol results in water-soluble and biocompatible polymeric nanoparticles, allowing for live cell molecular imaging at > 1000 nm with polymer fluorophores for the first time. Importantly, the high quantum yield of the polymer allows for *in vivo*, deep-tissue and ultrafast imaging of mouse arterial blood flow with an unprecedented frame rate of > 25 frames per second. The high time resolution results in spatially and time resolved imaging of the blood flow pattern in cardiogram waveform over a single cardiac cycle (~ 200 ms) of a mouse, which has not been observed with fluorescence imaging in this window before.

## Introduction

*In vivo* biological imaging in the second near-infrared window (NIR-II, 1.0-1.7 μm) has attracted much interest recently[1-8] owing to salient advantages over imaging in the visible (400-750 nm) and the conventional near-infrared (NIR, 750-900 nm) regions. Detecting longer wavelength photons in the NIR-II affords reduced photon scattering in biological tissues accompanied by lower autofluorescence, leading to higher spatial resolution at deeper tissue penetration depths.[9] To date, single-walled carbon nanotubes (SWNTs)[1-3,10-14], semiconducting quantum dots (QDs)[6,7,15-17], rare-earth doped nanoparticles[8] and nanoparticles made of small organic molecules[18] are fluorophores used for fluorescence imaging in the NIR-II region. These NIR-II agents can be functionalized with hydrophilic molecules to afford biocompatibility for intravenous administration and can be gradually cleared out from the body without obvious acute or chronic toxicity in animal studies.[16,19,20] However, shortcomings of existing NIR-II fluorophores include relatively low fluorescence quantum yield[1,8,21], unfavorably short emission wavelength[18] and potential toxicity due to heavy metals in QDs.[6,7,15-17] A much wider range of fluorophores could be developed to tackle these problems. An ideal fluorophore developed for NIR-II biological imaging should allow for tunable excitation and emission wavelengths in the > 1000 nm NIR-II region with a high fluorescence quantum yield and high biocompatibility.

Here we report the synthesis of conjugated polymers with intrinsic fluorescence > 1000 nm through donor-acceptor (D-A) alternating copolymerization, an effective way of synthesizing polymers with tunable bandgap energy in the NIR and at even longer wavelength regions.[22-29] By copolymerization of an electron-donating monomer benzo[1,2-b:3,4-b']difuran and an electron-withdrawing monomer fluorothieno-[3,4-b]thiophene, we derive a brightly fluorescent copolymer, poly(benzo[1,2-b:3,4-b']difuran-*alt*-fluorothieno-[3,4-b]thiophene) (named 'pDA') with fluorescence emission at ~1050 nm with a large Stokes shift of ~ 400 nm and a high fluorescence quantum yield of ~1.7%, much higher than the quantum yield of typical carbon nanotubes (~0.4%) used previously.[3] We non-covalently functionalize the polymer with a PEGylated surfactant to afford water solubility and biocompatibility. We successfully tune the emission wavelength of the conjugated polymer fluorophores from 1050 nm to 1350 nm, and demonstrate the first *in vitro* and *in vivo* biological imaging in the > 1000 nm window using conjugated polymer fluorophores.

## Results

**Synthesis of NIR-II Fluorescent pDA-PEG Nanoparticles.** The pDA polymer was synthesized through a copolymerization reaction of two monomers, 1-(4,6-dibromo-3-fluorothieno[3,4-b]thiophen-2-yl)nonan-1-one and 2, 6-bis(trimethyltin)-4,8-bis (2-ethylhexyloxy)benzo[1,2-b;3,4-b']difuran.[26] Fluorine was introduced into the thienothiophene unit to tune the fluorescence to long wavelengths (**Fig. 1**). The pDA polymer was characterized by $^1$H NMR spectroscopy (see Methods and **Supplementary Figs. 1-4**) and gel permeation chromatography (GPC). The average molecular weight ($M_n$) of the polymer was ~16 kDa (see Methods and **Supplementary Fig. 5**). We dissolved the as-made pDA polymer in tetrahydrofuran (THF) and then mixed it with an aqueous solution of a biocompatible surfactant of PEGylated phospholipid [DSPE-mPEG (5 kDa)][1,3,19] (see Methods). The mixture was dialyzed against water to remove THF, resulting in a stable aqueous solution of supramolecular conjugate comprised of hydrophobic core of pDA and hydrophilic shell of DSPE-mPEG coating (**Fig. 2a**; named as pDA-PEG).

Atomic force microscopy (AFM) revealed the size distribution of pDA-PEG nanoparticles (**Fig. 2b**) with an average size of 2.9 nm when dried on a silicon surface (**Supplementary Fig. 6**). The UV-Vis-NIR absorption spectrum of the as-made pDA-PEG solution in water exhibited an absorption peak at 654 nm, while the fluorescence emission spectrum showed a main emission peak at 1047 nm, suggesting a large Stokes shift of ~ 400 nm (**Fig. 2c**). The pDA copolymer has been reported to have red-shifted absorption and emission compared to the corresponding homopolymers, due to the formation of a charge-transfer structure between the electron donor and acceptor units.[26,27,30] The fluorescence quantum yield of pDA-PEG in a solution (**Fig. 2d**) under an excitation of 808 nm was ~ 1.7%, measured against a standard IR-26 dye as a reference[31,32] (**Supplementary Fig. 7a-f**, where the excitation of 808 nm was intentionally chosen to balance absorption and scattering to afford maximum penetration depth of excitation light for *in vivo* imaging). The quantum yield was much higher than that of typical nanotubes (~0.4%) (**Supplementary Fig. 7g-i**). The pDA-PEG exhibited high photostability in phosphate buffer saline (PBS) and fetal bovine serum with negligible decay under continuous excitation for 1 h (**Fig. 2e-f** and **Supplementary Fig. 8**). These results suggested pDA-PEG as an aqueous soluble, photo-stable and high brightness NIR-II fluorescent agent suited for biological imaging.

**Molecular Imaging of Cells with pDA-PEG in NIR-II.** We investigated pDA-PEG as a fluorescent label capable of targeting specific molecules on cell surfaces for performing molecular imaging of cancer cells through functionalization of pDA with targeting ligands. We made pDA-PEG-NH$_2$ using an amine-terminated phospholipid surfactant, DSPE-PEG-NH$_2$.[33,34] Cetuximab (Erbitux) antibodies were then thiolated and conjugated to pDA-PEG-NH$_2$ via standard crosslinking reaction between the -NH$_2$ groups on the polymer and -SH groups on the thiolated antibody (see Methods). Then the pDA-PEG-Erbitux was applied to target the epidermal growth factor receptors (EGFR) on the cell membranes of EGFR-positive breast tumor MDA-MB-468 cells while the EGFR-negative brain tumor U87-MG cells were used as a negative control.[15] Cell imaging in NIR-II detected > 1000 nm fluorescence of pDA-PEG-Erbitux (**Fig. 3**) selectively on the EGFR-positive MDA-MB-468 cells and not on the negative U87-MG cells, showing a positive/negative ratio of ~5.8. Thus, the pDA-PEG was made into an NIR-II fluorescent agent capable of recognizing and staining live cells with molecular specificity. This opened up the possibility of molecular imaging with conjugated copolymers *in vitro* and *in vivo*, which would not be attainable by structural imaging techniques such as ultrasound and optical coherence tomography (OCT).[35,36]

**Ultrafast Blood Flow Tracking with pDA-PEG in NIR-II.** Next, we performed *in vivo* mouse blood vessel imaging by detecting the > 1000 nm fluorescence of intravenously injected pDA-PEG solutions. Owing to the bright fluorescence of pDA-PEG with a 4-fold higher quantum yield (1.7%) than nanotube fluorophores[3] (quantum yield ~0.4%), we were able to dynamically image and track real-time arterial blood flow in the mouse hindlimb with a much shorter exposure time (~20 ms) than previously possible (~ 100 ms) in the NIR-II window.[3] Immediately following a tail-vein injection of 200 μL solution of pDA-PEG, we performed video-rate imaging of the left hindlimb of a mouse in the supine position with an ultrafast frame rate of 25.6 frames per second (fps) compared to the previously achievable 5.3 fps.[3] Such a high imaging speed enabled clear observation of the fast-moving flow front in the femoral artery as the NIR-II emitting pDA-PEG entered the hindlimb (**Supplementary Movie 1** and **Fig. 4a**). The average femoral blood flow velocity was quantified by plotting the distance traveled by the flow front as a function of time, showing an overall linear increase with an average blood velocity of 4.36 cm s$^{-1}$ (**Fig. 4b**), consistent with measurement by ultrasound (4~7 cm s$^{-1}$).[3] This represented the first *in vivo* blood flow velocity measurement by direct tracking through real-space imaging of fast-moving flow front on the millisecond scale using NIR-II fluorescence.

Interestingly, high time-resolution analysis of blood flow-front position versus time revealed periodic variations in instantaneous velocity over time with a period of 150-200 ms, showing oscillations between the highest instantaneous blood velocity of ~8 cm s$^{-1}$ and the lowest blood velocity of ~2 cm s$^{-1}$, corresponding to the ventricular ejection (systolic) and ventricular relaxation (diastolic) phases respectively of a single cardiac cycle (**Fig. 4c**). The observation of such a rapid periodic oscillation in the femoral blood velocity was entirely owed to the high time resolution of 20 ms, much shorter than the cardiac cycle of ~ 200 ms. The fast arterial blood flow (average speed ~ 4.36 cm s$^{-1}$) passed through the entire femoral artery (length ~2 cm) within two cardiac cycles (~400 ms). At longer times post injection, the same hindlimb imaged at 39 s post injection (p.i.) showed full perfusion of the injected NIR-II pDA-PEG fluorescent polymer into the femoral artery (**Fig. 4d**).

To further glean the oscillating blood flow front with high time-resolution, we selected a region of interest (ROI) of the femoral artery and analyzed the NIR-II fluorescence intensity as a function of time (**Fig. 5a**). We observed evenly spaced intensity humps over a linearly increasing baseline (**Fig. 5b**), with the humps corresponding to the systolic phases of cardiac cycles of the mouse and the increasing baseline due to increased overall perfusion of pDA-PEG fluorophore into the artery vessel.[3] By subtracting the linear rise from the intensity vs. time curve, we observed clear consecutive intensity spikes in the time course due to periodic ventricular ejections (**Fig. 5c**). Further, the linear-baseline subtracted fluorescence intensity vs. time was made into a movie that clearly showed the blood flow variations over the two phases of each heartbeat (**Figs. 5d-j** & **Supplementary Movie S2**). The blood flow vs. time showed a similar cardiogram waveform (**Fig. 5c**) as previous Doppler ultrasound measurement,[37] demonstrating fluorescence imaging could help visualize rapid blood flow changes in real-space and real-time within a single cardiac cycle of mouse.

It is also interesting to note that the dots observed along the femoral artery in the baseline-subtracted images (**Figs. 5f-h)** were at fixed positions over time and corresponded to local higher NIR-II intensity regions along the femoral artery. These high intensity local regions were found to coincide with locations along the vessel with slightly larger vessel diameters (by ~20%) of the lumen measured with the vessel image. That is, more NIR-II pDA-PEG fluorophores in the blood were filled into a larger lumen and thus exhibited higher NIR-II intensity (**Supplementary Fig. 9**). From the NIR-II fluorescence oscillations with an average period of 206.7 ms per cardiac cycle (**Fig. 5k**), we measured a heart rate of 290 beats per minute for the mouse, which agreed with previous results via cardiac gating.[38] Thus, the high-temporal and high-spatial resolution NIR-II imaging afforded by the high brightness of the conjugated polymer led to remarkably rich details of blood flows and cardiac cycles *in vivo*.

## Discussion

*In vivo* fluorescence imaging of live animals in the NIR-II region benefits from deeper penetration of up to a few millimeters inside the body and reduced scattering of photons that scales inversely proportional with wavelength as $\sim\lambda^{-w}$ ($w = 0.22\text{-}1.68$) in biological tissues.[3,9] Conjugated polymers have been widely used in organic solar cells,[26,39] light-emitting diodes[40,41] and organic electronics.[42,43] Fluorescent imaging with conjugated polymers has been limited to emission wavelength < 900 nm.[44-46] Our current work developed the NIR-II agent with > 1000 nm fluorescence based on conjugated polymers for biological imaging both *in vitro* and *in vivo*.[47] The facile synthetic route of donor-acceptor copolymers could allow further tuning of the optical properties of the polymers through modifying the donor-acceptor structures (**Supplementary Fig. 10**) and the length of the copolymer. A more electron-donating donor and a more electron-withdrawing acceptor typically result in smaller band gap of the copolymer,[48] and a longer copolymer exhibits reduced band gap due to a greater delocalization of π-electrons.[30] This could lead to a library of polymers with tunable excitation and emission wavelengths for NIR-II imaging (see **Supplementary Fig. 10** for several polymers synthesized thus far). With such development, we envisage fluorophores with different emission wavelengths in the 1.0-1.7 μm NIR-II window, allowing for multicolor molecular imaging of different biomarkers. Also, further improved penetration depth could be achieved *in vivo* by tuning the emission wavelength.

A salient advantage of the pDA-PEG polymer as an NIR-II fluorophore for *in vivo* live animal imaging is its much higher quantum yield (~1.7%) than SWNTs (~0.4%) previously used for hindlimb blood flow tracking.[3] This allows for much faster (20 ms exposure time vs. previous 100 ms)[3] video-rate imaging of dynamic changes in the blood flow labeled by NIR-II fluorophore, pushing the limit of temporal resolution to a previously unattainable level of > 25 fps (the instrument frame rate limit is 50 fps due to an overhead time of 19 ms). Although the pDA-PEG polymer has similar quantum efficiency (~1.7%) to nanoparticles of small organic molecules (~1.8%),[18] since the major peak of the small organic molecule nanoparticles is located at ~920 nm while the major peak of pDA-PEG polymer is at ~1050 nm, the pDA-PEG polymer has a greater portion of emitted photons in the >1000 nm NIR-II region (~70%) than the small molecule nanoparticles (~40%).[18] Taken together, the high fluorescence quantum yield, red-shifted fluorescence emission in the NIR-II region and low density of the pDA polymer require 10~100× lower mass dose of injection than previous NIR-II fluorophores (such as SWNTs[1-3,9-12], QDs[6,7,15-17], rare-earth doped nanoparticles[8] and nanoparticles incorporating organic molecules[18]) to reach the same *in vivo* imaging quality, and allow for dynamic fluorescence imaging with much shorter exposure time and higher temporal resolution. This high temporal resolution well exceeds the heart rate of mice (~5 beats per second, or ~ 200 ms per cycle) by 5 times, eliminating the need for cardiac gating typically required for low-frame-rate cardiovascular imaging techniques.[38,49]

Previous fluorescence imaging and tracking of blood flow usually require the removal of scattering tissue over the vessels of interest to afford higher spatial resolution, and gating devices to eliminate image blurring due to cardiac and respiratory motions on a faster time scale than the temporal resolution of image acquisitions (~10 fps).[38,50] Ultrasound imaging and optical coherence tomography (OCT) with ultrafast image acquisition rate (kHz~MHz) have shown the advantages of measuring fast blood flow and cardiac cycles,[51-53] but the spatial resolution (in the 10 μm ~ 1 mm range) and signal-to-noise ratio are sub-optimal due to long wavelengths[54] and speckles.[55,56] Here, by utilizing NIR-II fluorescence with reduced tissue scattering and an ultrafast frame rate of >25 fps, we have enabled simultaneous blood velocity and cardiac cycle measurements as well as high resolution imaging of blood vessels 1-2 mm deep under otherwise highly scattering skin tissues.[57] We have also proved the concept that the polymer-based NIR-II fluorophores can be used to track blood flow in capillaries with sub-10 μm diameters (**Supplementary Fig. 11a**), which are well below the spatial resolutions of traditional ultrasound and OCT. Based on the capillary blood flow tracking images, we measured the capillary blood velocity to be ~55.2 μm s$^{-1}$ (**Supplementary Fig. 11b**), in good agreement with previous studies.[58,59] The dynamic range of blood velocity measurement in our NIR-II imaging system using the pDA fluorophores is derived as 0 to 640 mm/s based on the brightness of the fluorophore and the detector sensitivity (**Supplementary Note 1**). In addition, we have shown that the pDA-PEG fluorophores can be used to track regional blood flow and redistribution as a result of increased metabolic demand after heat-induced inflammation of the tissue (**Supplementary Fig. 12**), providing a direct diagnostic tool for visualizing the metabolic difference of the tissue.

The heavy-metal-free nature of fluorescent conjugated polymers bodes well for potentially low toxicity agents for *in vivo* applications.[60] In a cellular toxicity study, we found that the toxicity of pDA-PEG nanoparticles depended on the surfactant coating, and a surfactant with branched PEG exhibited alleviated toxicity (see Methods). The mice injected with pDA-PEG were monitored over a period of up to 2 months without showing obvious toxic effects or health problems. We also attempted tuning the size of the pDA-PEG nanoparticles and obtained the smallest pDA-PEG with an average hydrodynamic size of < 6 nm (**Supplementary Fig. 13**, and also see Methods for more information). Systematic investigation is required to study the long term fate and toxicology of pDA-PEG nanoparticles, which is beyond the scope of the current work.

Real-time hemodynamic imaging could be of central importance to better understanding various cardiovascular diseases and designing treatment strategies.[61] An ideal NIR-II fluorescent agent capable of achieving hemodynamic imaging with high spatial and temporal resolutions should have tunable emission wavelength in the >1000 nm region to minimize tissue scattering, high fluorescence quantum efficiency (> 5%) for ultra-short imaging exposure time. The current work opens up future research to achieve these goals, and could eventually lead to NIR-II fluorescence agents suitable for clinic use.

## Methods

**Synthesis of pDA copolymer.** The synthesis of monomers M1 and M2 can be found in previous publications.[62-64] For the synthesis of the pDA copolymer, 0.091 g of monomer M1 (0.2 mmol), 0.148 g of the monomer M2 (0.2 mmol) and 15 mL of anhydrous toluene were mixed in a two-neck flask. The solution was flushed with $N_2$ for 10 min, and 15 mg of $Pd(PPh_3)_4$ was added into the flask. Then the solution was flushed with $N_2$ for another 25 min. The two-neck flask was placed in an oil bath and heated up carefully to 110 ℃. The mixture was stirred for 24 h at 110 ℃ under $N_2$ atmosphere. After the reaction was complete, the mixture was allowed to cool down to room temperature, and the polymer was precipitated by adding 100 mL of methanol and filtered through a Soxhlet thimble. Soxhlet extraction was performed with methanol, hexanes and chloroform. 80 mg of the polymer was obtained as a green solid from the chloroform fraction by rotary evaporation, and dried under vacuum overnight with a moderate yield of 56%.

**Characterizations of monomers and pDA copolymer.** All compounds, including the monomers M1 and M2, and the pDA copolymer, were characterized by nuclear magnetic resonance (NMR) spectroscopy. The $^1$H NMR and $^{19}$F NMR spectra were measured on a Bruker AV-400 spectrometer at room temperature. Chemical shifts were described as δ values (ppm), where tetramethylsilane (TMS) was used as an internal reference. Splitting patterns were labeled as s (singlet), d (doublet), t (triplet), m (multiplet) and br (broad). As such, the NMR spectral assignments for the two monomers M1 and M2, as well as the product polymer, are given as follows,

Monomer M1:

$^1$H NMR (400 MHz, $CDCl_3$): δ 2.95 (t, 2H), 1.76 (m, 2H), 1.39-1.30 (m, 10H), 0.91 (t, 3H).

$^{19}$F NMR (400 MHz, $CDCl_3$): -129 ppm (s, Ar-F)

Monomer M2:

$^1$H NMR (400 MHz, $CDCl_3$): δ 7.06 (s, 2H), 4.35 (d, 4H), 1.78 (m, 2H), 1.37-1.70 (m, 16 H), 0.96-1.03 (m, 12 H), 0.44 (s, 18 H).

Polymer pDA:

$^1$H NMR (400 MHz, CDCl$_3$): δ 6.80 (br, 2H), 4.31 (br, 4H), 3.06 (br, 2H), 2.01-1.21 (br, 30H), 0.81-1.21 (br, 15H).

Elemental analysis of the pDA copolymer was performed on a Flash EA 1112 elemental analyzer. Elemental composition of pDA copolymer was calculated based on the molecular formula (C$_{41}$H$_{53}$FO$_5$S$_2$)$_n$ as C: 69.26%; H: 7.80%; O: 11.25%, and compared to measured elemental composition by the elemental analyzer: C, 70.08%; H, 7.73%; O, 11.09%.

Molecular weight and polydispersity of the pDA copolymer were measured by gel permeation chromatography (GPC). The number-average molecular weights ($M_n$), weight-average molecular weights ($M_w$) and polydispersity index (PDI, $M_w/M_n$) were determined using polystyrene as the reference. Measurements were carried out using a Waters 515 HPLC pump, a Waters 2414 differential refractometer, and three Waters Styragel columns (HT2, HT3 and HT4) with THF as the eluent at a flow rate of 1.0 mL min$^{-1}$ at a temperature of 35 °C.

**Preparation of pDA-PEG and pDA-PEG-NH$_2$.** The as-synthesized pDA polymer was dissolved in THF at a mass concentration of 75 μg mL$^{-1}$ and mixed with an aqueous solution of DSPE-mPEG(5 kDa) (1,2-distearoyl-*sn*-glycero-3-phosphoethanolamine-N-[methoxy(polyethyleneglycol, 5000)], Laysan Bio) at a concentration of 1.1 mg mL$^{-1}$ with a 1:9 volume ratio for making the pDA-PEG, or with an aqueous solution of DSPE-PEG-NH$_2$ (5 kDa) (1,2-distearoyl-*sn*-glycero-3-phosphoethanolamine-N-[amino(polyethyleneglycol, 5000)], Laysan Bio) at 1.1 mg mL$^{-1}$ with a 1:9 volume ratio for making the pDA-PEG-NH$_2$. The pDA polymer remained soluble in the mixed solvent without any precipitation. Then the mixture was dialyzed against water to remove THF and make a THF-free, clear aqueous solution of pDA-PEG or pDA-PEG-NH$_2$ polymers. To remove aggregates formed during dialysis, the suspension was ultracentrifuged for 30 min at 300,000 g and only the supernatant was retained. Free, unbound surfactant in the solution was removed through 30-kDa centrifugal filters (Amicon) without causing any instability to the pDA-PEG polymers, suggesting the hydrophobic core of pDA and the hydrophilic shell of DSPE-mPEG were making a strongly-held complex.

**Preparation of pDA-PEG-Erbitux bioconjugate.** A 200 μL solution of pDA-PEG-NH$_2$ after excess surfactant removal was concentrated to a mass concentration of 320 μg mL$^{-1}$, determined by a mass extinction coefficient of 30.4 L g$^{-1}$ cm$^{-1}$ at its peak absorbance in the UV-Vis-NIR absorption spectrum. Then this solution was mixed with 1 mM sulfo-SMCC for 1 h in PBS at pH 7.4. On the other hand, 12.3 µL of Erbitux solution (Bristol-Myers Squibb) at a concentration of 2 mg mL$^{-1}$ (~13.3 µM) was mixed with a 1.64 µL solution of Traut's reagent (Sigma) at 1 mM concentration and 86 µL PBS solution. In the mixture the molar ratio of Erbitux antibody to Traut's reagent was 1:10. The mixture of Erbitux antibody and Traut's reagent was allowed to react for 1.5 h at room temperature. After removing excess sulfo-SMCC from the polymer solution and excess Traut's reagent from the antibody solution by filtration through 30-kDa centrifugal filters (Amicon), the two solutions were mixed and allowed to react for 2 days in PBS at 4 °C to make the pDA-PEG-Erbitux bioconjugate.

**Spectral Charaterizations of pDA-PEG.** UV-Vis-NIR absorption spectrum of pDA-PEG polymer in water was measured by a Cary 6000i UV-Vis-NIR spectrophotometer, background-corrected for contribution from water. The measured range was 400-900 nm. The NIR-II fluorescence spectrum was taken using a home-built NIR-II spectroscopy setup in the 900-1500 nm region. The excitation source was an 808-nm diode laser (RMPC lasers) at an output power of 160 mW and filtered by an 850-nm short-pass filter (Thorlabs), a 1000-nm short-pass filter (Thorlabs), an 1100-nm short-pass filter (Omega) and a 1300-nm short-pass filter (Omega). The excitation at 808 nm was intentionally chosen to balance absorption and scattering to afford maximum penetration depth of excitation light for *in vivo* imaging. The excitation beam was allowed to pass through an aqueous solution of pDA-PEG polymer in a 1 mm path cuvette (Starna Cells) and the emission was collected in the transmission geometry with a 900-nm long-pass filter (Thorlabs) to reject the excitation light. The emitted fluorescence from the sample was directed into a spectrometer (Acton SP2300i) equipped with a liquid-nitrogen-cooled InGaAs linear array detector (Princeton OMA-V). The emission spectrum was corrected after raw data acquisition to account for the detector sensitivity profile and extinction feature of the filter using the MATLAB software.

**Determination of fluorescence quantum yield of pDA-PEG.** Fluorescence quantum yield of the pDA-PEG polymer in water was measured in a similar way to a previous publication,[32] using the NIR-II fluorescent IR-26 dye as the reference (quantum yield = 0.5%).[31] For reference calibration on our setup, a stock solution of 1 mg mL$^{-1}$ IR-26 dissolved in 1,2-dichloroethane (DCE) was diluted to a DCE solution with its absorbance value of ~0.10 at 808 nm. Then this sample was diluted using DCE to a series of samples with their absorbance values at 808 nm of ~0.08, ~0.06, ~0.04 and ~0.02. Using these highly diluted samples with low absorbance minimized any secondary optical processes such as reabsorption and reemission effects. Then a total of five IR-26 solutions in DCE with linearly spaced concentrations (including the one with absorbance of ~0.10) were loaded into a 10-mm path fluorescence cuvette (Starna Cells) one at a time. The excitation source was an 808-nm diode laser (RMPC lasers) with a linewidth FWHM of 1.89 nm at an output power of 160 mW and filtered by an 850-nm short-pass filter (Thorlabs), a 1000-nm short-pass filter (Thorlabs), an 1100-nm short-pass filter (Omega) and a 1300-nm short-pass filter (Omega). The emission was collected in the transmission geometry with a 900-nm long-pass filter (Thorlabs) to reject the excitation light and the emission spectrum was taken in the 900-1500 nm region. The same absorption and emission measurements were carried out for pDA-PEG polymers and SWNTs in aqueous solutions too. Then all emission spectra of both the reference and the samples were corrected after raw data acquisition to account for the detector sensitivity profile and the extinction profiles of the filters, and integrated in the 900-1500 nm NIR-II region. The definition of quantum yield dictates that all emitted photons should be taken into account for quantum yield measurement.[65] In our quantum yield measurements, since an 808-nm laser is used, ideally the integration of the emission spectra should start from 809 nm to cover the complete emission band; however, due to the limited choice of excitation and emission filters, we started the integration from 900 nm, which would give an underestimated quantum yield. Nonetheless, this underestimation would not be too much since one could see from the emission spectra that the emission dropped rapidly below 950 nm, meaning there would not be too much fluorescence emission under 900 nm. The integrated NIR-II fluorescence intensity was plotted against absorbance at the excitation wavelength of 808 nm and fitted into a linear function. Two slopes, one obtained from the reference of IR-26 in DCE and the other from the sample (pDA-PEG or SWNT), were employed in the calculation of the quantum yield of the sample, based on equation (1) as follows,

$$QY_{\text{sample}} = QY_{\text{ref}} \cdot \frac{slope_{\text{sample}}}{slope_{\text{ref}}} \cdot \left(\frac{n_{\text{sample}}}{n_{\text{ref}}}\right)^2 \qquad (1)$$

where $n_{\text{sample}}$ and $n_{\text{ref}}$ are the refractive indices of water and DCE, respectively.

**AFM imaging of pDA-PEG.** AFM image of pDA-PEG was acquired with a Nanoscope IIIa multimode AFM in the tapping mode. The sample for imaging was prepared by drop-drying a highly diluted aqueous solution of pDA-PEG polymer at a mass concentration of 750 ng $L^{-1}$ onto the $SiO_2$/silicon substrate without any post-processing steps. To plot the size distribution of the pDA-PEG polymeric nanoparticles, 100 nanoparticles were measured from the AFM images and their heights were used to generate the histogram of size distribution. Due to the effect of tip size convolution in AFM,[66] the height measurement from the AFM micrograph, rather than the lateral size measurement, was used to measure the size of the nanoparticles deposited on the substrate.

**Dynamic light scattering (DLS) analysis of pDA-PEG.** DLS was performed using a solution of pDA-PEG at 0.75 μg $ml^{-1}$ in PBS in a 1-cm quartz cuvette. A PBS solution was used to mimic the ionic strength in blood because the hydrodynamic size of pDA-PEG was dependent on the ionic strength. The data was collected using a Brookhaven Instruments 90Plus Particle Size Analyzer.

**Cell incubation and staining.** All cell culture media were supplemented with 10% fetal bovine serum, 100 μg·$mL^{-1}$ streptomycin, 100 IU $mL^{-1}$ penicillin and L-glutamine. The human glioblastoma U87-MG cells were cultured in Low Glucose Dulbecco's Modified Eagle Medium (DMEM), with 1 g $L^{-1}$ D-glucose and 110 mg $L^{-1}$ sodium pyruvate, in the presence of 5% $CO_2$. Human breast cancer MDA-MB-468 cells were cultured in Leibovitz's L-15 medium at 37 ℃, $CO_2$ free. Cells were maintained in a 37 ℃ incubator. For cell staining, cells were trypsinized before the as-made pDA-PEG-Erbitux bioconjugate was incubated with both EGFR-positive (MDA-MB-468) and EGFR-negative (U87-MG) cell lines at a concentration of 5 μg $mL^{-1}$ at 4 ℃ for 1 h, followed by washing the cells with 1x PBS to remove all free, unbound pDA-PEG-Erbitux bioconjugate in the suspension.

**NIR-II fluorescence imaging of cancer cells.** Targeted cell imaging in the 1.0-1.7 μm NIR-II window was carried out using a 658-nm laser diode with a 150 μm-diameter excitation beam focused by a ×50 objective lens (Olympus). The NIR-II fluorescence from pDA-PEG-Erbitux bioconjugate was collected using a liquid-nitrogen-cooled, 320 × 256 pixel, 2D InGaAs camera (Princeton Instruments 2D OMA-V). The excitation light was filtered out using a 900 nm and a 1000 nm long-pass filter (Thorlabs) so that the intensity of each pixel in the camera represented fluorescence in the 1.0-1.7 μm NIR-II region. The exposure time was 3 s for all fluorescence images.

**Mouse Handling.** All vertebrate animal experiments were performed under the approval of Stanford University's Administrative Panel on Laboratory Animal Care (APLAC). 8-week old female BALB/c mice were obtained from Charles River and housed at the Research Animal Facility of Stanford under our approved animal protocols. Before hindlimb vessel imaging, all mice were anesthesized in a rodent anesthesia machine with 2 L min$^{-1}$ O$_2$ gas flow mixed with 3% Isoflurane. Then the hair over the hindlimb skin was carefully removed using Nair to avoid causing wounds to the skin. Tail vein injection of the pDA-PEG contrast agent was carried out in dark and synchronized with the camera that started continuous image acquisition simultaneously. The injected dose was a 200 μL bolus of the pDA-PEG in a 1×PBS solution at a mass concentration of 0.25 mg mL$^{-1}$. During the time course of imaging the mouse was kept anesthetized by a nose cone delivering 1.5 L min$^{-1}$ O$_2$ gas mixed with 3% Isoflurane.

**Ultrafast video-rate NIR-II vessel imaging.** Video-rate NIR-II vessel imaging was done in a similar manner as our previous publication.[3] In brief, imaging was carried out on a homebuilt imaging setup consisting of a 2D InGaAs camera (Princeton Instruments 2D OMA-V). The excitation was provided by an optical-fiber-coupled 808-nm diode laser (RMPC Lasers), intentionally chosen to balance absorption and scattering to afford maximum penetration depth of excitation light for *in vivo* imaging. The high fluorescence quantum yield of the pDA-PEG also ensured sufficient emission for ultra-fast *in vivo* imaging despite the non-resonant excitation under 808 nm. The light was collimated by a collimator with a focal length of 4.5 mm (ThorLabs) and filtered by an 850-nm and a 1000-nm shortpass filter (ThorLabs). The power density of the excitation laser at the imaging plane was 140 mW cm$^{-2}$, significantly lower than the reported safe exposure limit of 329 mW cm$^{-2}$ at 808 nm.[67] The emitted fluorescence from the mouse was allowed to pass through a 900 nm and a 1000 nm long-pass filter (Thorlabs) and to be focused onto the InGaAs detector by a lens pair consisting of two NIR achromats (200 and 75 mm; Thorlabs). The distance between the two NIR achromats was adjusted to have only one hindlimb of the mouse included in the field of view. The camera was set to expose continuously using the LabVIEW software with an exposure time of 20 ms. NIR-II images were acquired with a frame rate of 25.6 fps due to a 19 ms overhead time during readout.

***In vitro* toxicity study of pDA-PEG.** We determined the *in vitro* toxicity of pDA-PEG by MTS assay using a CellTiter96 kit (Promega) on human breast cancer MDA-MB-468 cells. To evaluate how the surfactant coating of the pDA polymer affected the nanoparticles' *in vitro* toxicity, we coated the pDA polymer (the same formulation as in **Fig. 1**) with DSPE-mPEG and poly(maleic anhydride-alt-1-octadecene)methoxy(polyethyleneglycol, 5kDa) (C18-PMH-mPEG), respectively, where DSPE-mPEG has one linear PEG chain while C18-PMH-mPEG has many branched PEG chains. For *in vitro* toxicity study, approximately 5,000 MDA-MB-468 cells were incubated per well with 100 μl of Leibovitz's L-15 medium spiked with pDA-DSPE-mPEG or pDA-C18-PMH-mPEG at serially diluted concentrations ($n = 6$ for each concentration). The cells were kept at 37 °C in a humidified atmosphere for 36 h in the absence of $CO_2$. 100% MDA-MB-468 cell medium without any pDA-PEG was taken as the positive control while MDA-MB-468 cell medium spiked with 20 μM Doxil® was taken as the negative control. Immediately before addition of 15 μl of CellTiter96 for the colorimetric assay, the polymer-spiked medium was removed from each well in the plate and replaced with fresh MDA-MB-468 cell growth medium. This would prevent any interference in the absorbance readings from the intrinsic color of pDA-PEG. After 1.5 h, the color change was quantified using a plate reader, which took absorbance readings at 485 nm. Cell viability was plotted as a fraction of the absorbance value of the positive control wells incubated without any pDA-PEG after baseline subtraction of the absorbance for medium alone. The negative control wells had an average cell viability value of ~60% after 36 h of incubation, which was as expected. We found from the *in vitro* toxicity study that pDA-DSPE-mPEG has a half maximal inhibitory concentration (IC50) level of ~30 μg/mL, while pDA-C18-PMH-mPEG has an IC50 level of >150 μg/mL. The pDA-PEG solution could not be further concentrated to an even higher concentration, preventing us from measuring the cell viability under an incubation concentration of >150 μg/mL. It is also noteworthy that the linear or branched PEG coating did not alter the fluorescence quantum efficiency to a noticeable level.

**Hydrodynamic size tuning of pDA-PEG nanoparticles.** The hydrodynamic size tuning was performed by coating the pDA polymer with surfactants of different PEG molecular weights (2 kDa and 5 kDa) and changing the initial pDA concentration in the THF solution (0.025-0.075 mg/mL; please see the "Preparation of pDA-PEG and pDA-PEG-$NH_2$" section in Methods) before mixing with 1.1 mg/mL DSPE-mPEG in water. After the pDA-PEG nanoparticles were made in each synthesis condition, dynamic light scattering (DLS) analysis was performed to evaluate each sample's hydrodynamic diameter as a function of initial pDA concentration in the THF solution and the molecular weight of PEG in the surfactant (please see the "Dynamic light scattering (DLS) analysis of pDA-PEG" section in Methods). The dependence of pDA-PEG hydrodynamic size on initial pDA concentration and PEG molecular weight is plotted in **Supplementary Fig. 13**, showing that both smaller PEG and lower initial concentration of pDA favor the formation of smaller pDA-PEG nanoparticles in aqueous solution. However, it is noteworthy that DLS was not able to discriminate the 'empty' nanoparticles without any pDA molecules loaded inside, and therefore the size distribution of the measured hydrodynamic diameters below ~4 nm could be due to the micelles formed by the DSPE-mPEG surfactants only.

# References

1	Welsher, K., Liu, Z., Sherlock, S. P., Robinson, J. T., Chen, Z., Daranciang, D. & Dai, H. J. A route to brightly fluorescent carbon nanotubes for near-infrared imaging in mice. *Nat. Nanotechnol.* **4**, 773-780 (2009).
2	Hong, G. S., Wu, J. Z., Robinson, J. T., Wang, H. L., Zhang, B. & Dai, H. J. Three-dimensional imaging of single nanotube molecule endocytosis on plasmonic substrates. *Nat. Commun.* **3**, 700 (2012).
3	Hong, G. S., Lee, J. C., Robinson, J. T., Raaz, U., Xie, L. M., Huang, N. F., Cooke, J. P. & Dai, H. J. Multifunctional in vivo vascular imaging using near-infrared II fluorescence. *Nat. Med.* **18**, 1841-1846 (2012).
4	Won, N., Jeong, S., Kim, K., Kwag, J., Park, J., Kim, S. G. & Kim, S. Imaging depths of near-infrared quantum dots in first and second optical windows. *Mol. Imaging* **11**, 338-352 (2012).
5	Yi, H. J., Ghosh, D., Ham, M. H., Qi, J. F., Barone, P. W., Strano, M. S. & Belcher, A. M. M13 Phage-Functionalized Single-Walled Carbon Nanotubes As Nanoprobes for Second Near-Infrared Window Fluorescence Imaging of Targeted Tumors. *Nano Lett.* **12**, 1176-1183 (2012).
6	Hong, G. S., Robinson, J. T., Zhang, Y. J., Diao, S., Antaris, A. L., Wang, Q. B. & Dai, H. J. In Vivo Fluorescence Imaging with Ag2S Quantum Dots in the Second Near-Infrared Region. *Angew. Chem. Int. Ed.* **51**, 9818-9821 (2012).
7	Dong, B. H., Li, C. Y., Chen, G. C., Zhang, Y. J., Zhang, Y., Deng, M. J. & Wang, Q. B. Facile Synthesis of Highly Photoluminescent Ag2Se Quantum Dots as a New Fluorescent Probe in the Second Near-Infrared Window for in Vivo Imaging. *Chem. Mater.* **25**, 2503-2509 (2013).
8	Naczynski, D. J., Tan, M. C., Zevon, M., Wall, B., Kohl, J., Kulesa, A., Chen, S., Roth, C. M., Riman, R. E. & Moghe, P. V. Rare-earth-doped biological composites as in vivo shortwave infrared reporters. *Nat. Commun.* **4**, 2199 (2013).
9	Welsher, K., Sherlock, S. P. & Dai, H. J. Deep-tissue anatomical imaging of mice using carbon nanotube fluorophores in the second near-infrared window. *P. Natl. Acad. Sci. USA* **108**, 8943-8948 (2011).
10	Robinson, J. T., Welsher, K., Tabakman, S. M., Sherlock, S. P., Wang, H. L., Luong, R. & Dai, H. J. High Performance In Vivo Near-IR (> 1 μm) Imaging and Photothermal Cancer Therapy with Carbon Nanotubes. *Nano Res.* **3**, 779-793 (2010).
11	Robinson, J. T., Hong, G. S., Liang, Y. Y., Zhang, B., Yaghi, O. K. & Dai, H. J. In Vivo Fluorescence Imaging in the Second Near-Infrared Window with Long Circulating Carbon Nanotubes Capable of Ultrahigh Tumor Uptake. *J. Am. Chem. Soc.* **134**, 10664-10669 (2012).
12	Diao, S., Hong, G. S., Robinson, J. T., Jiao, L. Y., Antaris, A. L., Wu, J. Z., Choi, C. L. & Dai, H. J. Chirality Enriched (12,1) and (11,3) Single-Walled Carbon Nanotubes for Biological Imaging. *J. Am. Chem. Soc.* **134**, 16971-16974 (2012).
13	Liu, Z., Tabakman, S., Sherlock, S., Li, X. L., Chen, Z., Jiang, K. L., Fan, S. S. & Dai, H. J. Multiplexed Five-Color Molecular Imaging of Cancer Cells and Tumor Tissues with Carbon Nanotube Raman Tags in the Near-Infrared. *Nano Res.* **3**, 222-233 (2010).
14	Liu, Z., Tabakman, S., Welsher, K. & Dai, H. J. Carbon Nanotubes in Biology and Medicine: In vitro and in vivo Detection, Imaging and Drug Delivery. *Nano Res.* **2**, 85-120 (2009).
15	Zhang, Y., Hong, G. S., Zhang, Y. J., Chen, G. C., Li, F., Dai, H. J. & Wang, Q. B. Ag2S Quantum Dot: A Bright and Biocompatible Fluorescent Nanoprobe in the Second Near-Infrared Window. *ACS Nano* **6**, 3695-3702 (2012).
16	Zhang, Y., Zhang, Y. J., Hong, G. S., He, W., Zhou, K., Yang, K., Li, F., Chen, G. C., Liu, Z., Dai, H. J. & Wang, Q. B. Biodistribution, pharmacokinetics and toxicology of Ag2S near-infrared quantum dots in mice. *Biomaterials* **34**, 3639-3646 (2013).

17    Zhu, C. N., Jiang, P., Zhang, Z. L., Zhu, D. L., Tian, Z. Q. & Pang, D. W. Ag2Se Quantum Dots with Tunable Emission in the Second Near-Infrared Window. *ACS Appl. Mater. Inter.* **5**, 1186-1189 (2013).
18    Tao, Z. M., Hong, G. S., Shinji, C., Chen, C. X., Diao, S., Antaris, A. L., Zhang, B., Zou, Y. P. & Dai, H. J. Biological Imaging Using Nanoparticles of Small Organic Molecules with Fluorescence Emission at Wavelengths Longer than 1,000 nm. *Angew. Chem. Int. Ed.* **52**, 13002-13006 (2013).
19    Liu, Z., Davis, C., Cai, W. B., He, L., Chen, X. Y. & Dai, H. J. Circulation and long-term fate of functionalized, biocompatible single-walled carbon nanotubes in mice probed by Raman spectroscopy. *P. Natl. Acad. Sci. USA* **105**, 1410-1415 (2008).
20    Schipper, M. L., Nakayama-Ratchford, N., Davis, C. R., Kam, N. W. S., Chu, P., Liu, Z., Sun, X. M., Dai, H. J. & Gambhir, S. S. A pilot toxicology study of single-walled carbon nanotubes in a small sample of mice. *Nat. Nanotechnol.* **3**, 216-221 (2008).
21    O'Connell, M. J., Bachilo, S. M., Huffman, C. B., Moore, V. C., Strano, M. S., Haroz, E. H., Rialon, K. L., Boul, P. J., Noon, W. H., Kittrell, C., Ma, J. P., Hauge, R. H., Weisman, R. B. & Smalley, R. E. Band gap fluorescence from individual single-walled carbon nanotubes. *Science* **297**, 593-596 (2002).
22    Zhang, Q. T. & Tour, J. M. Low optical bandgap polythiophenes by an alternating donor/acceptor repeat unit strategy. *J. Am. Chem. Soc.* **119**, 5065-5066 (1997).
23    Greenham, N. C., Moratti, S. C., Bradley, D. D. C., Friend, R. H. & Holmes, A. B. Efficient Light-Emitting-Diodes Based on Polymers with High Electron-Affinities. *Nature* **365**, 628-630 (1993).
24    Zhang, Z. G., Min, J., Zhang, S. Y., Zhang, J., Zhang, M. J. & Li, Y. F. Alkyl chain engineering on a dithieno[3,2-b:2 ',3 '-d]silole-alt-dithienylthiazolo[5,4-d]thiazole copolymer toward high performance bulk heterojunction solar cells. *Chem. Commun.* **47**, 9474-9476 (2011).
25    Cui, C. H., Fan, X., Zhang, M. J., Zhang, J., Min, J. & Li, Y. F. A D-A copolymer of dithienosilole and a new acceptor unit of naphtho[2,3-c]thiophene-4,9-dione for efficient polymer solar cells. *Chem. Commun.* **47**, 11345-11347 (2011).
26    Liu, B., Chen, X. W., Zou, Y. P., He, Y. H., Xiao, L., Xu, X. J., Li, L. D. & Li, Y. F. A benzo[1,2-b: 4,5-b ']difuran- and thieno-[3,4-b]thiophene-based low bandgap copolymer for photovoltaic applications. *Polym. Chem.* **4**, 470-476 (2013).
27    Yamamoto, T. *et al.* pi-conjugated donor-acceptor copolymers constituted of pi-excessive and pi-deficient arylene units. Optical and electrochemical properties in relation to CT structure of the polymer. *J. Am. Chem. Soc.* **118**, 10389-10399 (1996).
28    Huo, L. J., Ye, L., Wu, Y., Li, Z. J., Guo, X., Zhang, M. J., Zhang, S. Q. & Hou, J. H. Conjugated and Nonconjugated Substitution Effect on Photovoltaic Properties of Benzodifuran-Based Photovoltaic Polymers. *Macromolecules* **45**, 6923-6929 (2012).
29    Huo, L. J., Hou, J. H., Chen, H. Y., Zhang, S. Q., Jiang, Y., Chen, T. L. & Yang, Y. Bandgap and Molecular Level Control of the Low-Bandgap Polymers Based on 3,6-Dithiophen-2-yl-2,5-dihydropyrrolo[3,4-c]pyrrole-1,4-dione toward Highly Efficient Polymer Solar Cells. *Macromolecules* **42**, 6564-6571 (2009).
30    Cornil, J., Gueli, I., Dkhissi, A., Sancho-Garcia, J. C., Hennebicq, E., Calbert, J. P., Lemaur, V., Beljonne, D. & Bredas, J. L. Electronic and optical properties of polyfluorene and fluorene-based copolymers: A quantum-chemical characterization. *J. Chem. Phys.* **118**, 6615-6623 (2003).
31    Murphy, J. E., Beard, M. C., Norman, A. G., Ahrenkiel, S. P., Johnson, J. C., Yu, P. R., Micic, O. I., Ellingson, R. J. & Nozik, A. J. PbTe colloidal nanocrystals: Synthesis, characterization, and multiple exciton generation. *J. Am. Chem. Soc.* **128**, 3241-3247 (2006).
32    Williams, A. T. R., Winfield, S. A. & Miller, J. N. Relative Fluorescence Quantum Yields Using a Computer-Controlled Luminescence Spectrometer. *Analyst* **108**, 1067-1071 (1983).

33  Liu, Z., Tabakman, S. M., Chen, Z. & Dai, H. J. Preparation of carbon nanotube bioconjugates for biomedical applications. *Nat. Protoc.* **4**, 1372-1382 (2009).
34  Hong, G. S., Tabakman, S. M., Welsher, K., Chen, Z., Robinson, J. T., Wang, H. L., Zhang, B. & Dai, H. J. Near-Infrared-Fluorescence-Enhanced Molecular Imaging of Live Cells on Gold Substrates. *Angew. Chem. Int. Ed.* **50**, 4644-4648 (2011).
35  Riccabona, M., Nelson, T. R., Pretorius, D. H. & Davidson, T. E. Distance and Volume Measurement Using 3-Dimensional Ultrasonography. *Journal of Ultrasound in Medicine* **14**, 881-886 (1995).
36  Tomlins, P. H. & Wang, R. K. Theory, developments and applications of optical coherence tomography. *J. Phys. D Appl. Phys.* **38**, 2519-2535 (2005).
37  Wladimiroff, J. W., Tonge, H. M. & Stewart, P. A. Doppler Ultrasound Assessment of Cerebral Blood-Flow in the Human-Fetus. *Brit. J. Obstet. Gynaec.* **93**, 471-475 (1986).
38  Lee, S., Vinegoni, C., Feruglio, P. F., Fexon, L., Gorbatov, R., Pivoravov, M., Sbarbati, A., Nahrendorf, M. & Weissleder, R. Real-time in vivo imaging of the beating mouse heart at microscopic resolution. *Nat. Commun.* **3** (2012).
39  Li, G., Shrotriya, V., Huang, J. S., Yao, Y., Moriarty, T., Emery, K. & Yang, Y. High-efficiency solution processable polymer photovoltaic cells by self-organization of polymer blends. *Nat. Mater.* **4**, 864-868 (2005).
40  Burroughes, J. H., Bradley, D. D. C., Brown, A. R., Marks, R. N., Mackay, K., Friend, R. H., Burns, P. L. & Holmes, A. B. Light-Emitting-Diodes Based on Conjugated Polymers. *Nature* **347**, 539-541 (1990).
41  Kawamura, Y., Yanagida, S. & Forrest, S. R. Energy transfer in polymer electrophosphorescent light emitting devices with single and multiple doped luminescent layers. *J. Appl. Phys.* **92**, 87-93 (2002).
42  Facchetti, A. pi-Conjugated Polymers for Organic Electronics and Photovoltaic Cell Applications. *Chem. Mater.* **23**, 733-758 (2011).
43  Ong, B. S., Wu, Y. L., Liu, P. & Gardner, S. High-performance semiconducting polythiophenes for organic thin-film transistors. *J. Am. Chem. Soc.* **126**, 3378-3379 (2004).
44  Ghoroghchian, P. P., Frail, P. R., Susumu, K., Blessington, D., Brannan, A. K., Bates, F. S., Chance, B., Hammer, D. A. & Therien, M. J. Near-infrared-emissive polymersomes: Self-assembled soft matter for in vivo optical imaging. *P. Natl. Acad. Sci. USA* **102**, 2922-2927 (2005).
45  Jin, Y. H., Ye, F. M., Zeigler, M., Wu, C. F. & Chiu, D. T. Near-Infrared Fluorescent Dye-Doped Semiconducting Polymer Dots. *ACS Nano* **5**, 1468-1475 (2011).
46  Pu, K. Y., Li, K. & Liu, B. A Molecular Brush Approach to Enhance Quantum Yield and Suppress Nonspecific Interactions of Conjugated Polyelectrolyte for Targeted Far-Red/Near-Infrared Fluorescence Cell Imaging. *Adv. Funct. Mater.* **20**, 2770-2777 (2010).
47  Pu, K. Y. & Liu, B. Fluorescent Conjugated Polyelectrolytes for Bioimaging. *Adv. Funct. Mater.* **21**, 3408-3423 (2011).
48  Wang, Y. L., Peng, Q., Hou, Q. F., Zhao, K., Liang, Y. & Li, B. L. Tuning the electronic structures and optical properties of fluorene-based donor-acceptor copolymers by changing the acceptors: a theoretical study. *Theoretical Chemistry Accounts* **129**, 257-270 (2011).
49  Wiesmann, F., Szimtenings, M., Frydrychowicz, A., Illinger, R., Hunecke, A., Rommel, E., Neubauer, S. & Haase, A. High-resolution MRI with cardiac and respiratory gating allows for accurate in vivo atherosclerotic plaque visualization in the murine aortic arch. *Magnet. Reson. Med.* **50**, 69-74 (2003).
50  Gioux, S., Ashitate, Y., Hutteman, M. & Frangioni, J. V. Motion-gated acquisition for in vivo optical imaging. *J. Biomed. Opt.* **14** (2009).

51    Li, P., Yin, X., Shi, L., Rugonyi, S. & Wang, R. K. K. In vivo functional imaging of blood flow and wall strain rate in outflow tract of embryonic chick heart using ultrafast spectral domain optical coherence tomography. *J. Biomed. Opt.* **17** (2012).
52    An, L., Guan, G. Y. & Wang, R. K. K. High-speed 1310 nm-band spectral domain optical coherence tomography at 184,000 lines per second. *J. Biomed. Opt.* **16** (2011).
53    Gill, R. W. Measurement of Blood-Flow by Ultrasound - Accuracy and Sources of Error. *Ultrasound Med. Biol.* **11**, 625-641 (1985).
54    Xuan, J. W. *et al.* Functional neoangiogenesis imaging of genetically engineered mouse prostate cancer using three-dimensional power Doppler ultrasound. *Cancer Res.* **67**, 2830-2839 (2007).
55    Adler, D. C., Ko, T. H. & Fujimoto, J. G. Speckle reduction in optical coherence tomography images by use of a spatially adaptive wavelet filter. *Opt. Lett.* **29**, 2878-2880 (2004).
56    Abd-Elmoniem, K. Z., Youssef, A. B. M. & Kadah, Y. M. Real-time speckle reduction and coherence enhancement in ultrasound imaging via nonlinear anisotropic diffusion. *IEEE T. Bio.-Med. Eng.* **49**, 997-1014 (2002).
57    Hong, G., Lee, J. C., Jha, A., Diao, S., Nakayama, K. H., Hou, L., Doyle, T. C., Robinson, J. T., Antaris, A. L., Dai, H., Cooke, J. P. & Huang, N. F. Near-Infrared II Fluorescence for Imaging Hindlimb Vessel Regeneration with Dynamic Tissue Perfusion Measurement. *Circ. Cardiovasc. Imaging*, 10.1161/CIRCIMAGING.1113.000305 (2014).
58    Lin, W. C., Wu, C. C., Huang, T. C., Lin, W. C., Chiu, B. Y. C., Liu, R. S. & Lin, K. P. Red Blood Cell Velocity Measurement in Rodent Tumor Model: An in vivo Microscopic Study. *J. Med. Biol. Eng.* **32**, 97-102 (2012).
59    An, L., Qin, J. & Wang, R. K. Ultrahigh sensitive optical microangiography for in vivo imaging of microcirculations within human skin tissue beds. *Opt. Express* **18**, 8220-8228 (2010).
60    Tian, Z. Y., Yu, J. B., Wu, C. F., Szymanski, C. & McNeill, J. Amplified energy transfer in conjugated polymer nanoparticle tags and sensors. *Nanoscale* **2**, 1999-2011 (2010).
61    Schrier, R. W. & Abraham, W. T. Mechanisms of disease - Hormones and hemodynamics in heart failure. *N. Engl. J. Med.* **341**, 577-585 (1999).
62    Liang, Y. Y., Feng, D. Q., Wu, Y., Tsai, S. T., Li, G., Ray, C. & Yu, L. P. Highly Efficient Solar Cell Polymers Developed via Fine-Tuning of Structural and Electronic Properties. *J. Am. Chem. Soc.* **131**, 7792-7799 (2009).
63    Huo, L. J., Huang, Y., Fan, B. H., Guo, X., Jing, Y., Zhang, M. J., Li, Y. F. & Hou, J. H. Synthesis of a 4,8-dialkoxy-benzo[1,2-b:4,5-b ']difuran unit and its application in photovoltaic polymer. *Chem. Commun.* **48**, 3318-3320 (2012).
64    Chen, X. W., Liu, B., Zou, Y. P., Xiao, L., Guo, X. P., He, Y. H. & Li, Y. F. A new benzo[1,2-b:4,5-b ']difuran-based copolymer for efficient polymer solar cells. *J. Mater. Chem.* **22**, 17724-17731 (2012).
65    Wurth, C., Grabolle, M., Pauli, J., Spieles, M. & Resch-Genger, U. Relative and absolute determination of fluorescence quantum yields of transparent samples. *Nat. Protoc.* **8**, 1535-1550 (2013).
66    Keller, D. Reconstruction of Stm and Afm Images Distorted by Finite-Size Tips. *Surf. Sci.* **253**, 353-364 (1991).
67    Matthes, R. *et al.* Revision of guidelines on limits of exposure to laser radiation of wavelengths between 400 nm and 1.4 mu m. *Health Physics* **79**, 431-440 (2000).

## Acknowledgments

This study was supported by grants from the National Cancer Institute of US National Institute of Health to H. D. (5R01CA135109-02), grants from Natural Science Foundation of China (NSFC Nos. 51173206, 21161160443), National High Technology Research and Development Program (No. 2011AA050523) and Central South University to Y. Z., and a William S. Johnson Fellowship to G. H.

## Author contributions

H.D., G.H., Y.Z., and A.L.A. conceived and designed the experiments. G.H., Y.Z., A.L.A., S.D., D.W., K.C., X.Z., C.C., B.L., J.Z.W., J.Y., B.Z., Z.T. and C.F. performed the experiments. G.H., Y.Z., A.L.A., Y.H. and H.D. analyzed the data and wrote the manuscript. All authors discussed the results and commented on the manuscript.

## Additional information

This paper is now published online at *Nature Communications*.

**Figures**

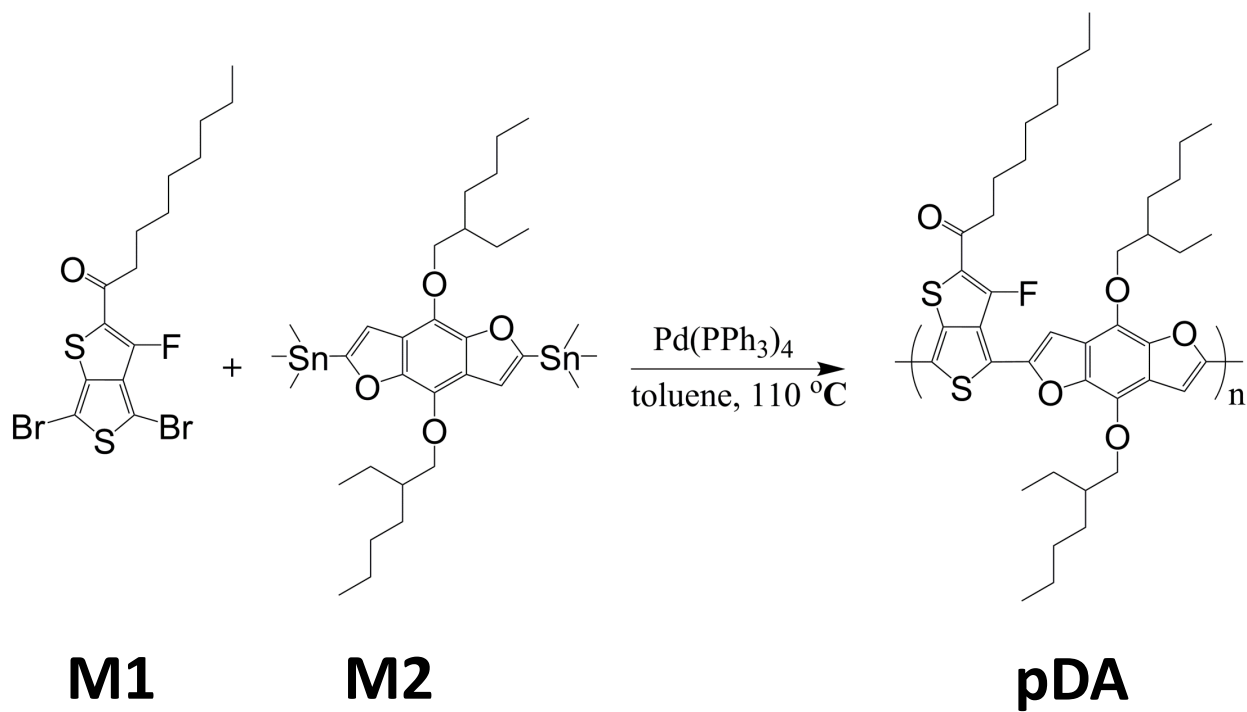

**Figure 1 | Scheme for the Synthesis of pDA Polymer.** This scheme shows the chemical structures of the two monomers M1 and M2, along with the structure of the pDA polymer, poly(benzo[1,2-b:3,4-b']difuran-*alt*-fluorothieno-[3,4-b]thiophene).

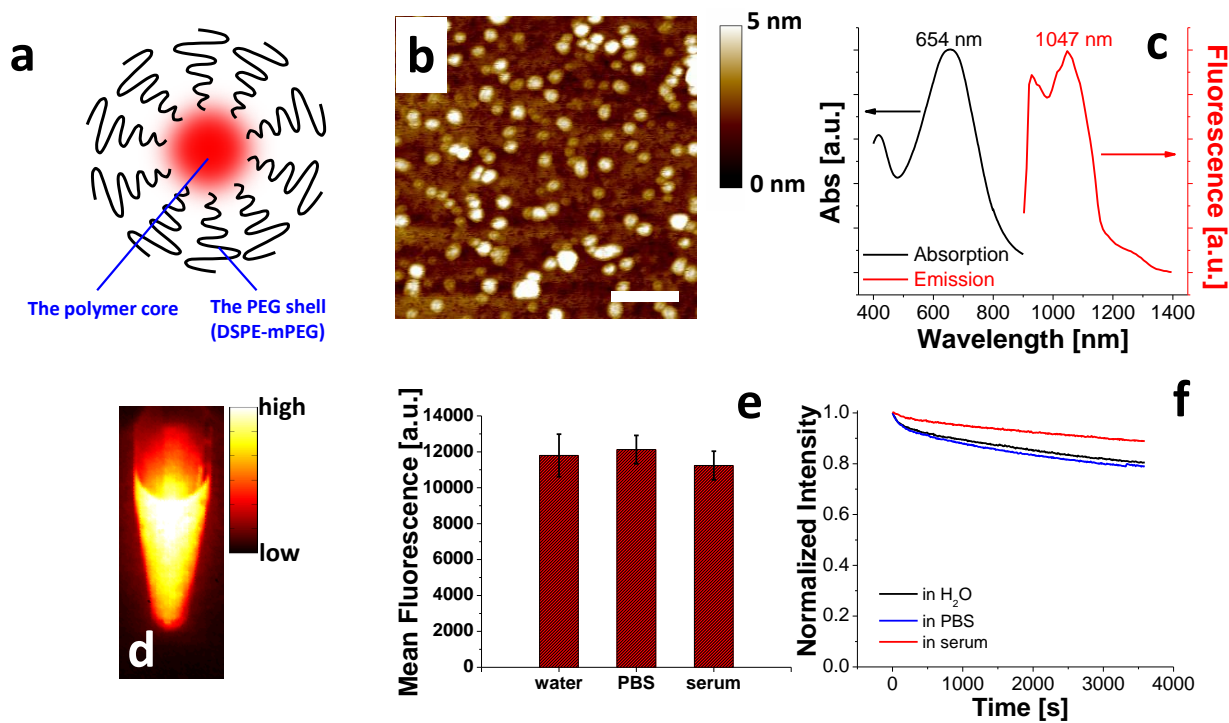

**Figure 2 | Characterizations of pDA-PEG nanoparticle.** (**a**) A schematic of the pDA-PEG nanoparticle showing a hydrophobic polymer core and hydrophilic PEG shell. (**b**) A typical AFM image of pDA-PEG nanoparticles deposited on a silicon substrate. Due to the effect of tip size convolution in AFM,[66] the height measurement from the AFM micrograph, rather than the lateral size measurement, was used to measure the size of the nanoparticles deposited on the substrate. The lateral scale bar in **b** indicates 100 nm. (**c**) Absorption and emission spectra of pDA-PEG, featuring a large Stokes shift of ~ 400 nm. (**d**) An NIR-II fluorescence image of an aqueous solution of pDA-PEG taken in the range of 1.0-1.7 μm NIR-II window under an excitation of 808 nm. (**e**) Fluorescence stability of pDA-PEG in different media including water, PBS and serum. The error bars were obtained by taking the standard deviation of all pixel intensities within an ROI covering the solution in the 1-mm cuvette, as shown in **Supplementary Fig. 8**. (**f**) Photostability curves of pDA-PEG in water, PBS and serum under continuous 808-nm illumination. pDA-PEG in serum exhibits the lowest degree of photobleaching (<10%) among the three.

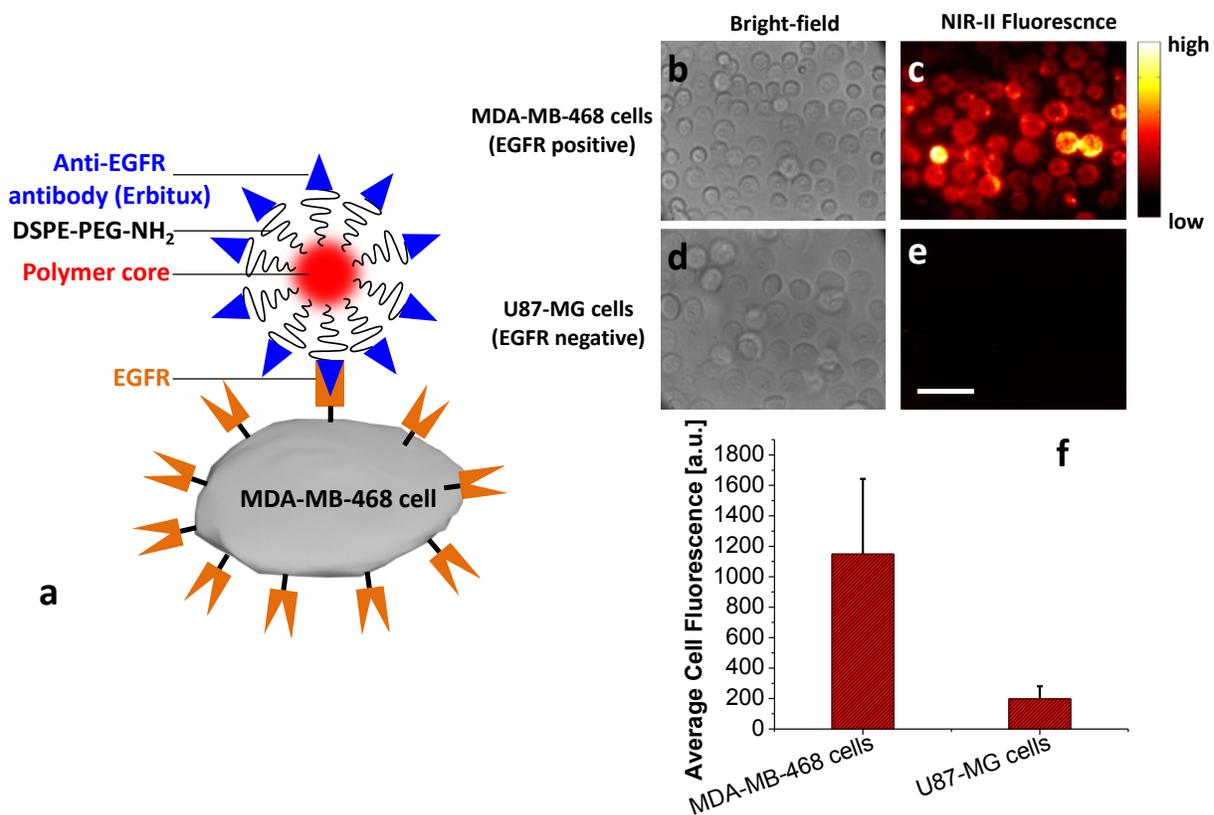

**Figure 3 | Molecular Cell Imaging with pDA-PEG-Erbitux.** (**a**) A schematic showing the structure of pDA-PEG-Erbitux bioconjugate, where the anti-EGFR antibody (Erbitux) selectively targets EGFR on the cell membrane of an MDA-MB-468 cell. (**b**&**c**) White-light (**b**) and NIR-II (**c**) fluorescence images of EGFR positive MDA-MB-468 cells incubated with the pDA-PEG-Erbitux bioconjugate, showing positive staining of cells. (**d**&**e**) White-light (**d**) and NIR-II (**e**) fluorescence images of EGFR negative U87-MG cells incubated with the pDA-PEG-Erbitux bioconjugate, without obvious staining of the cells. The scale bar in **e** indicates 40 μm, which applies to all images shown in **b**-**e**. (**f**) Average NIR-II fluorescence of EGFR positive MDA-MB-468 cells and negative U87-MG cells, showing a positive/negative ratio of ~5.8. The error bars in **f** were obtained by taking the standard deviation of average fluorescence intensity from 20 cells in each NIR-II fluorescence image.

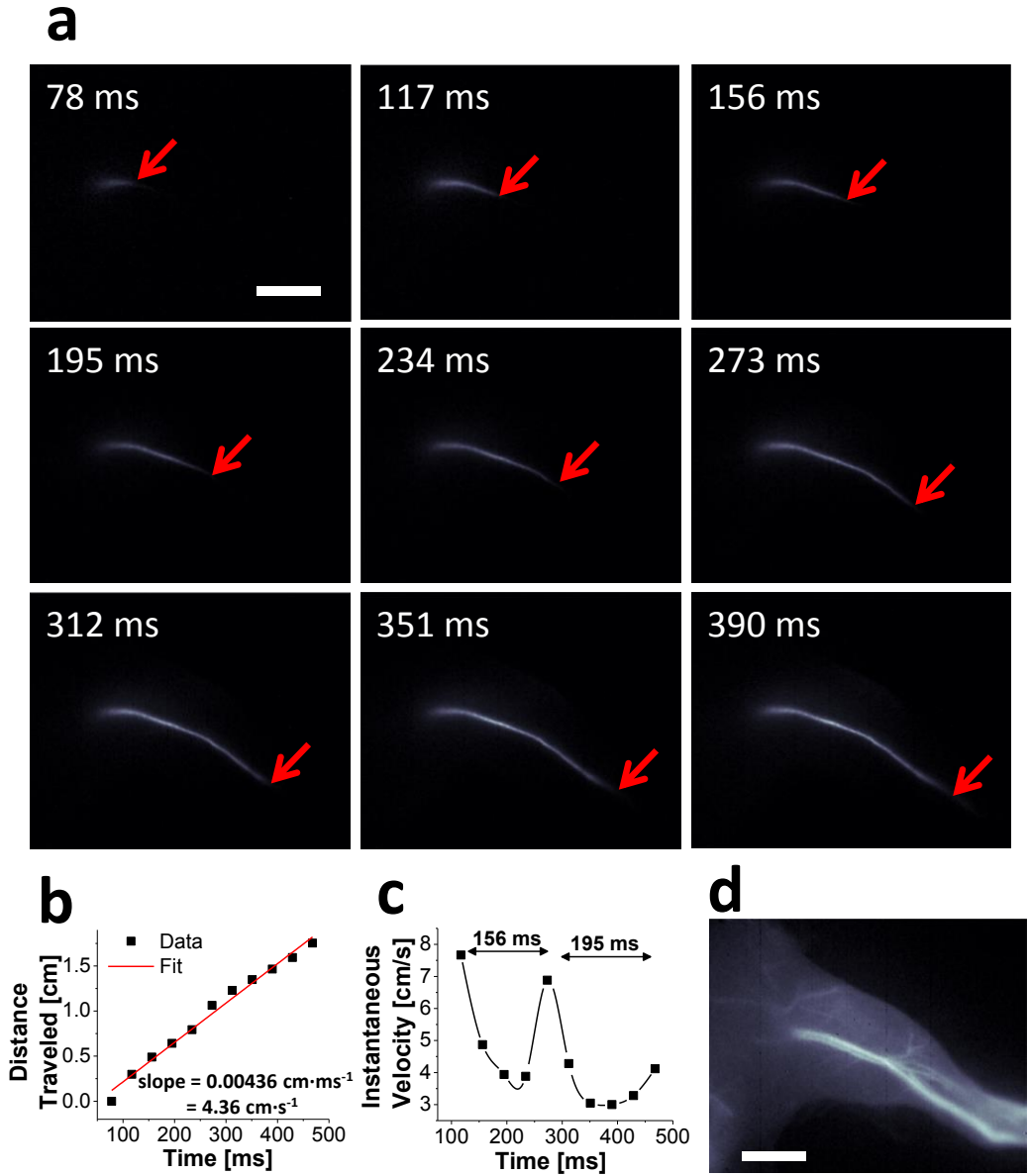

**Figure 4 | Ultrafast NIR-II Imaging of Arterial Blood Flow.** (**a**) A time course of NIR-II fluorescence images of a mouse hindlimb immediately following intravenous injection of pDA-PEG, showing the blood flow front moving inside the femoral artery (indicated by red arrows). The frame rate of imaging is 25.6 fps with an exposure time of 20 ms and an instrument overhead time of 19 ms. (**b**) A plot of the distance travelled by the blood flow front as a function of time. The linear fit reveals an average blood velocity of 4.36 cm s$^{-1}$ in the femoral artery. (**c**) A plot of instantaneous velocity (derived by dividing flow front traveled distance between two consecutive frames by the time interval of 39 ms) as a function of time, revealing periodic changes of instantaneous velocity corresponding to cardiac cycles. (**d**) An NIR-II fluorescence image of the same mouse hindlimb after full perfusion of pDA-PEG-containing blood into the hindlimb, upon which the fluorescence intensity in the hindlimb became unchanging. The scale bars in **a**&**d** indicate 5 mm.

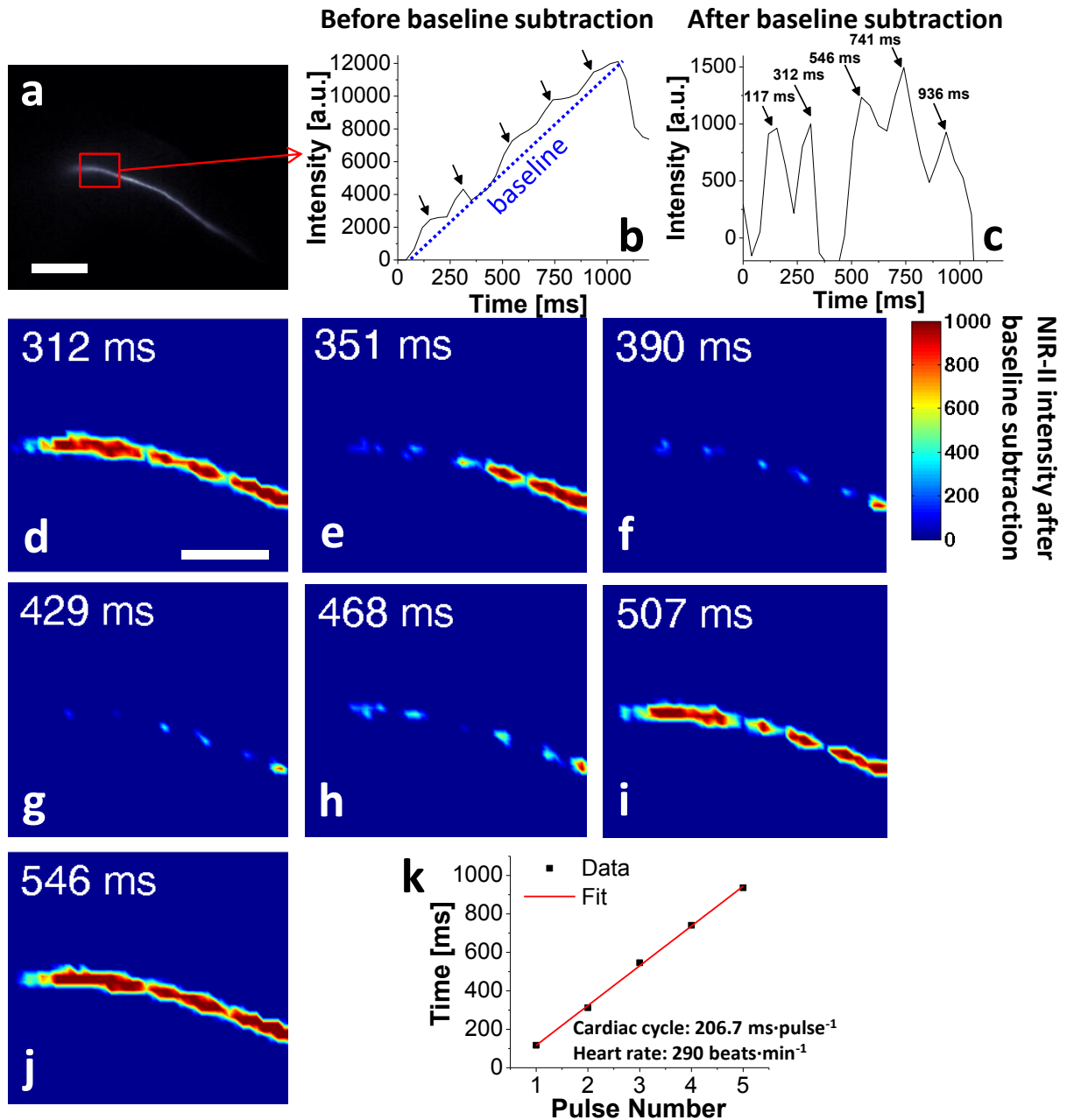

**Figure 5 | Resolving Waveform Blood Flow Pattern with pDA.** (**a&b**) An NIR-II fluorescence image (**a**) of the mouse femoral artery, where the fluorescence intensity inside the region of interest (ROI) red box is integrated and plotted as a function of time in (**b**), showing an increasing profile with humps corresponding to ventricular ejections of cardiac cycles. The scale bar in **a** indicates 5 mm. (**c**) NIR-II fluorescence intensity plotted as a function of time, after a linear increasing baseline is subtracted from the plot shown in **b**, featuring 5 cardiac cycles in the plot. (**d-j**) Time course NIR-II fluorescence images of the red box area shown in **a**, after

subtraction of a time-dependent linearly-increasing background given by the baseline in **b**. Note that these seven images correspond to a complete cardiac cycle from 312 ms to 546 ms. See **Supplementary Movie 2** for a video showing the real-time evolution of the linear-background subtracted fluorescence intensity. The scale bar in **d** indicates 1 mm, which applies to all images of **d**-**j**.(**k**) Time point of NIR-II fluorescence spikes corresponding to ventricular ejections shown in **c**, plotted over several heart pulses (black squares). The data is fitted to a linear function with its slope of 206.7 ms per pulse corresponding to the period of each cardiac cycle.

# Supplementary Figures

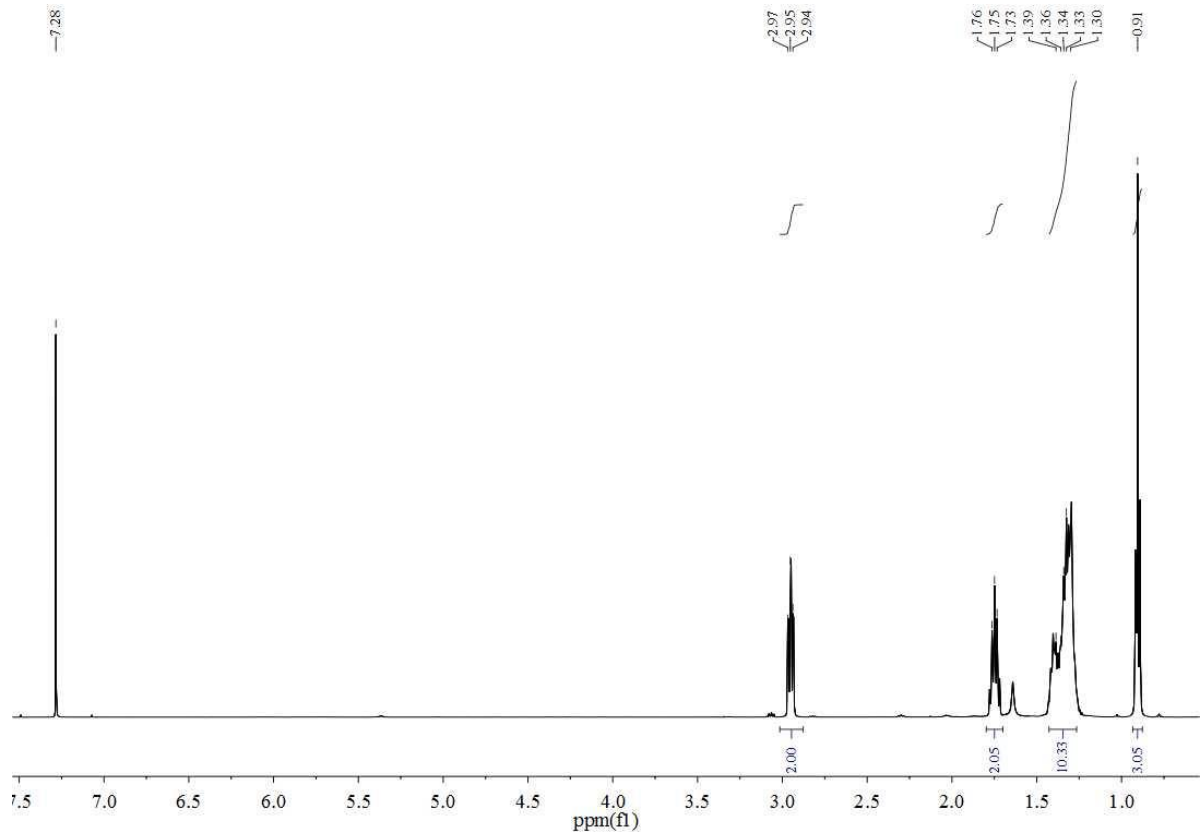

**Supplementary Figure 1 | $^1$H NMR Spectrum of the Monomer M1.** The peaks in the NMR spectrum are assigned as follows: 2.95 (t, 2H), 1.76 (m, 2H), 1.39-1.30 (m, 10H), 0.91 (t, 3H).

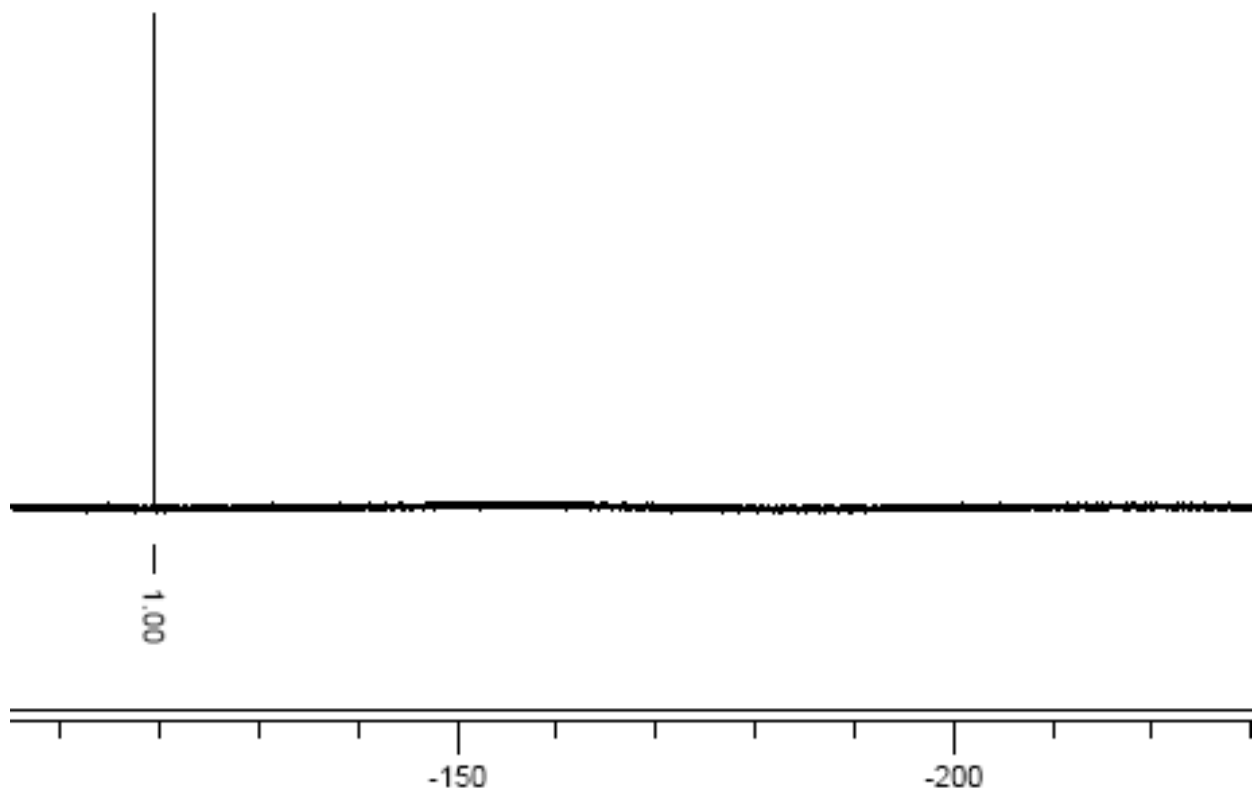

**Supplementary Figure 2 | $^{19}$F NMR Spectrum of the Monomer M1.** The peak in the $^{19}$F NMR spectrum is assigned as follows: -129 ppm (s, Ar-F)

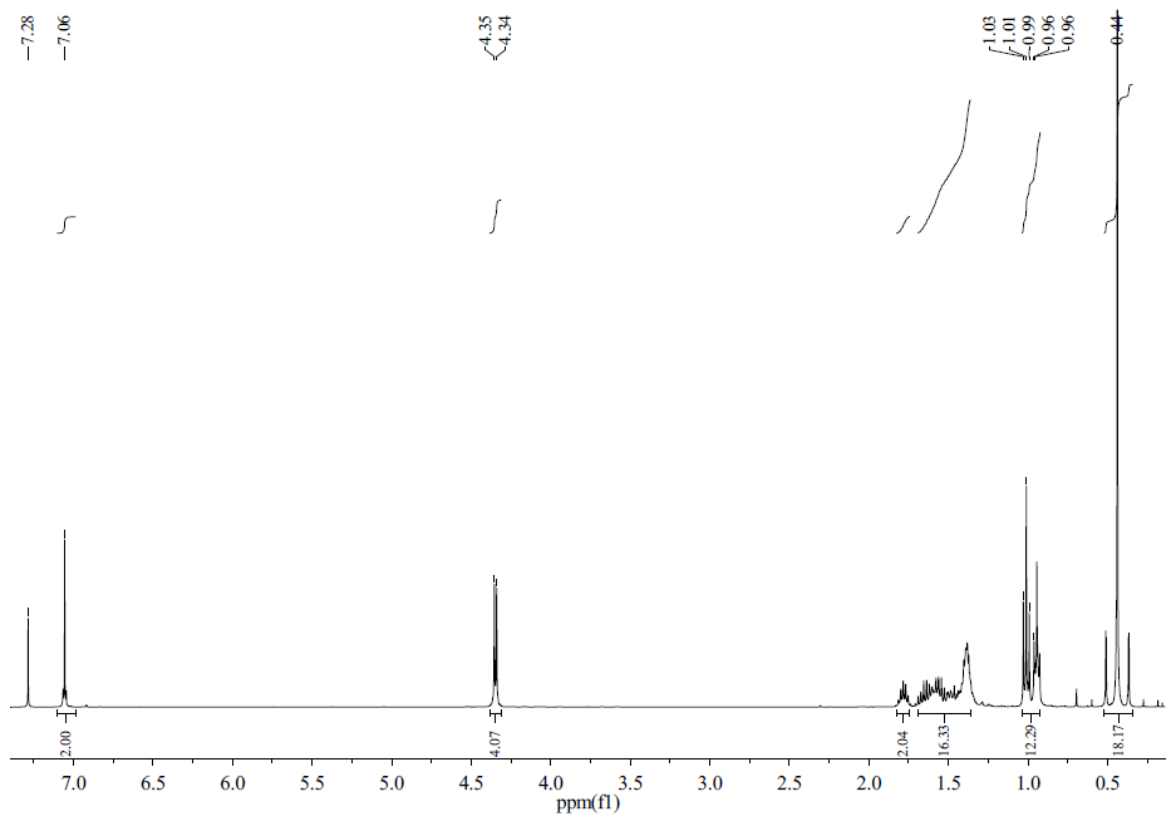

**Supplementary Figure 3 | $^1$H NMR Spectrum of the Monomer M2.** The peaks in the NMR spectrum are assigned as follows: 7.06 (s, 2H), 4.35 (d, 4H), 1.78 (m, 2H), 1.37-1.70 (m, 16 H), 0.96-1.03 (m, 12 H), 0.44 (s, 18 H).

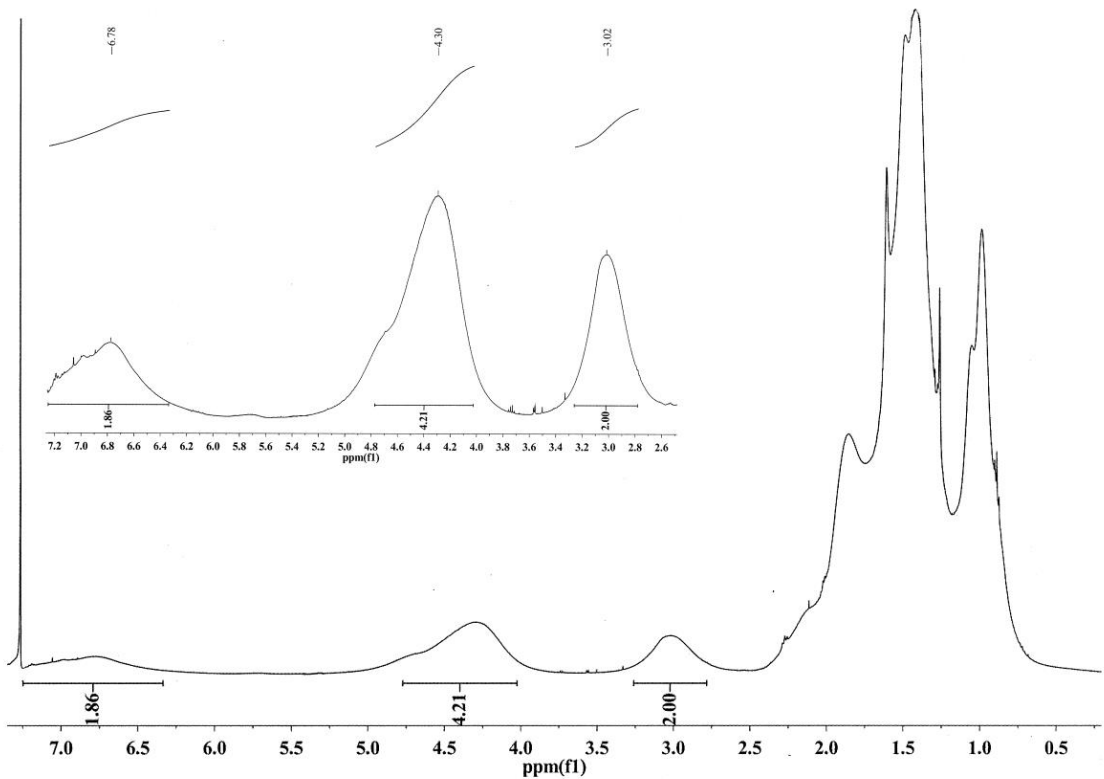

**Supplementary Figure 4 | $^1$H NMR Spectrum of the pDA Polymer.** The peaks in the NMR spectrum are assigned as follows: 6.80 (br, 2H), 4.31 (br, 4H), 3.06 (br, 2H), 2.01-1.21 (br, 30H), 0.81-1.21 (br, 15H). The inset shows a zoomed-in view of the spectrum in the 2.5-7.3 ppm region.

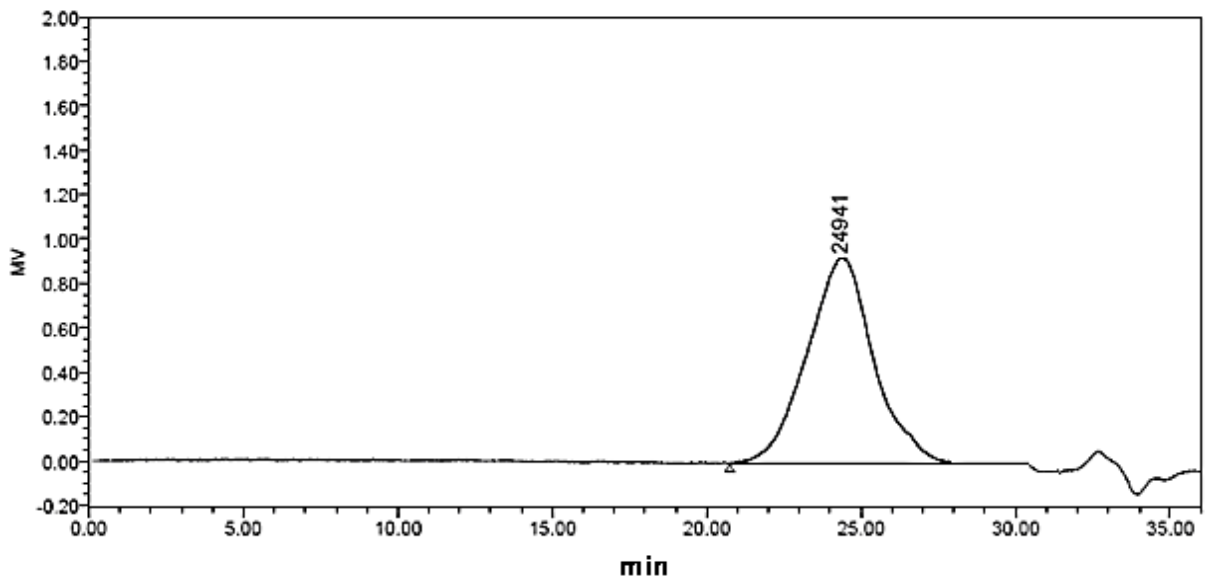

**Supplementary Figure 5 | Gel Permeation Chromatography (GPC) Spectrum of pDA.**
Molecular weight and polydispersity of the polymer are listed as follows: $M_n$ (number-average molecular weight): 16,192; $M_w$ (weight-average molecular weight): 30,991; $M_p$ (peak molecular weight): 24,941; PDI (polydispersity index):1.91.

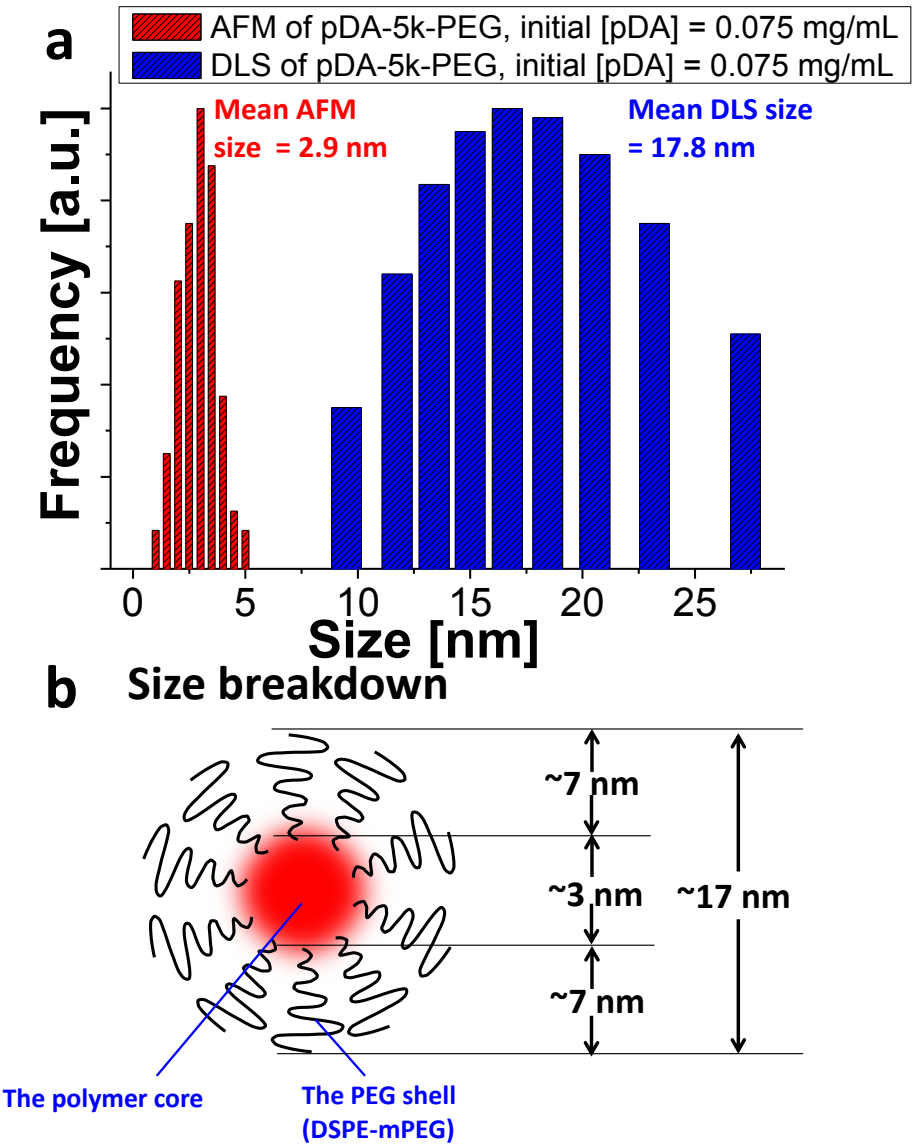

**Supplementary Figure 6 | Size Analysis of pDA-PEG.** (**a**) A bar chart showing the particle size histogram of dried pDA-5k-PEG nanoparticles from the height measurements of the AFM image shown in **Fig. 2b** (red bars), and the size distribution of the same pDA-5k-PEG sample in a 1x PBS solution based on DLS measurement (blue bars). (**b**) The size breakdown of the pDA-5k-PEG complex showing the polymer core of ~3 nm (measured by AFM for the collapsed nanoparticle) and the overall size of ~17 nm when the PEG chains were hydrated and extended (measured by DLS in an aqueous solution).

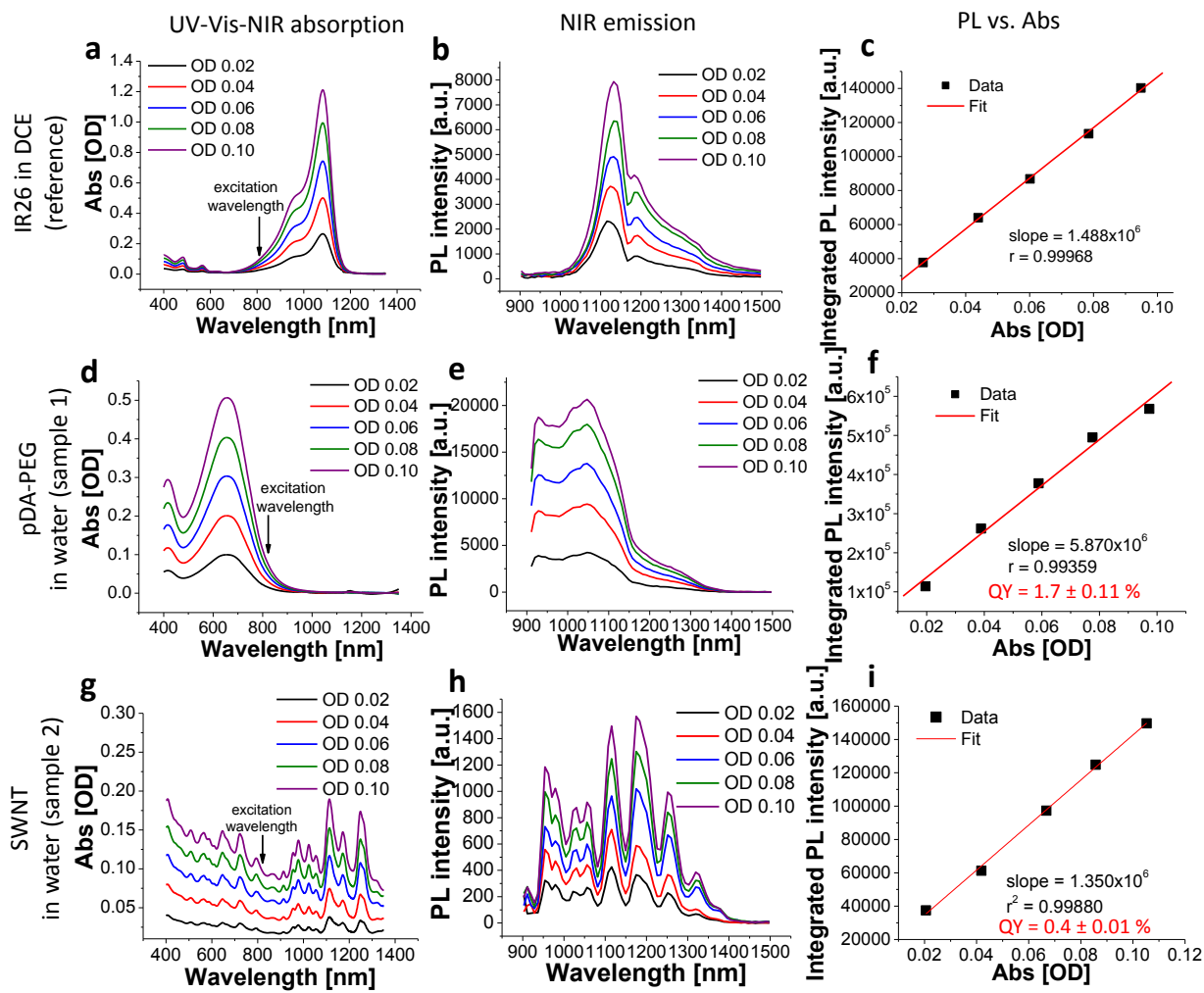

**Supplementary Figure 7 | Quantum Yield Measurements.** (**a**) UV-Vis-NIR absorption spectra of a series of the IR-26 reference solutions in DCE with increasing concentrations. (**b**) NIR-II emission spectra of the IR-26 reference solutions shown in **a** under an excitation of 808 nm. (**c**) Integrated NIR-II fluorescence intensity plotted as a function of absorbance at 808 nm for IR-26 reference solutions based on the measurements in **a** and **b**. The data was fitted into a linear function with a slope of $1.488 \times 10^6$. (**d**) UV-Vis-NIR absorption spectra of a series of the pDA-PEG polymer solutions in water with increasing concentrations. (**e**) NIR-II emission spectra of the pDA-PEG polymer solutions shown in **d** under an excitation of 808 nm. (**f**) Integrated NIR-II fluorescence intensity plotted as a function of absorbance at 808 nm for pDA-PEG polymer solutions based on the measurements in **d** and **e**. The data was fitted into a linear function with a slope of $5.870 \times 10^6$, giving a measured quantum yield of $1.7 \pm 0.11\%$. (**g**) UV-Vis-NIR absorption spectra of a series of SWNT solutions in water with increasing concentrations. (**h**) NIR-II emission spectra of the SWNT solutions shown in **g** under an excitation of 808 nm. (**i**) Integrated NIR-II fluorescence intensity plotted as a function of absorbance at 808 nm for SWNT

solutions based on the measurements in **g** and **h**. The data was fitted into a linear function with a slope of $1.350 \times 10^6$, giving a measured quantum yield of $0.4 \pm 0.01\%$.

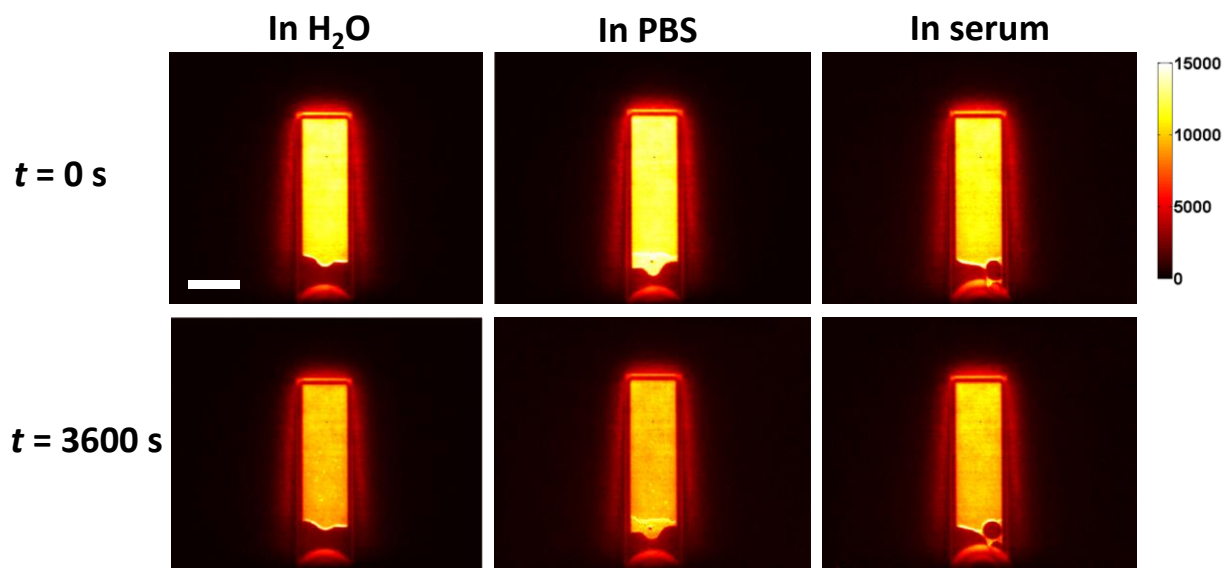

**Supplementary Figure 8 | Photostability of the pDA-PEG Polymer in Different Media.** The top row shows the NIR-II fluorescence images of pDA-PEG polymer in water, PBS and serum before continuous 808-nm excitation, at the same concentration of 7.5 μg/mL. The bottom row shows the NIR-II fluorescence images of the same samples after continuous 808-nm excitation for 1 h. The scale bar indicates 1 cm.

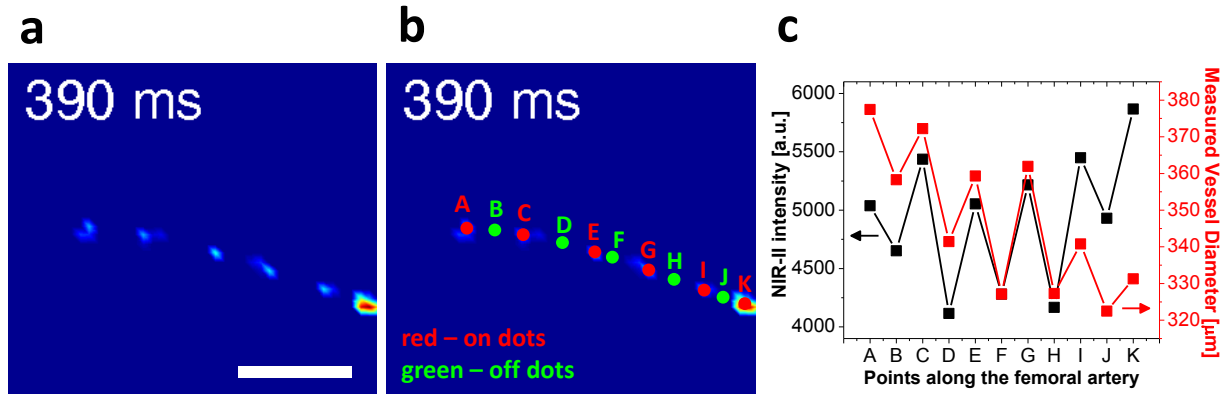

**Supplementary Figure 9 | Origin of 'Dot' Features in Fig. 5.** (**a**) An NIR-II fluorescence image at 390 ms p.i., corresponding to the red box area shown in **Fig. 5a**, after subtraction of the baseline shown in **Fig. 5b**. The scale bar indicates 1 mm. (**b**) The same basedline-subtracted NIR-II image as in **a**, with all 'dot' features labeled as A, C, E, G, I and K (red), and the locations between two neighboring dots labeled as B, D, F, H and J (green). (**c**) NIR-II fluorescence intensity and vessel diameter measured at all points labeled from A to K in **b**, showing in-phase changes between the two measurements at these points. A positive correlation between the local blood volume and the NIR-II fluorescence intensity has been found, suggesting a larger lumen (i.e., a larger diameter) of the vessel could hold more blood with more NIR-II fluorescent agents, and thus appeared brighter than neighboring vessel segments, resulting in the 'dot' features after subtracting a constant baseline from all pixels in the NIR-II fluorescence image.

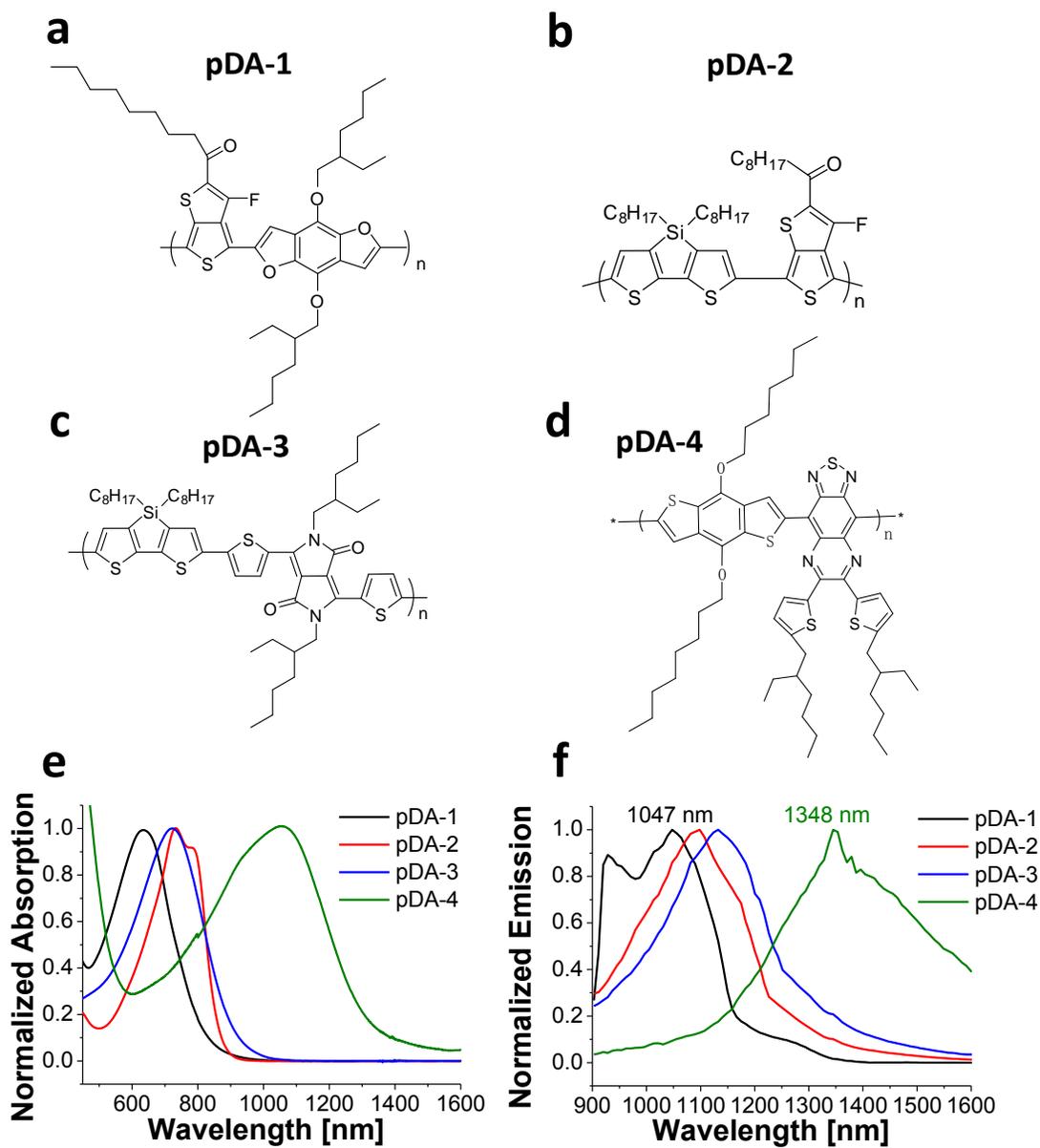

**Supplementary Figure 10 | pDA Polymers with Tunable Absorption and Emission Properties.** The chemical structures of four different pDA molecules are shown in **a**-**d**, along with their absorption and emission spectra (**e&f**) revealing the emission wavelengths tuned in the range of 1050-1350 nm.

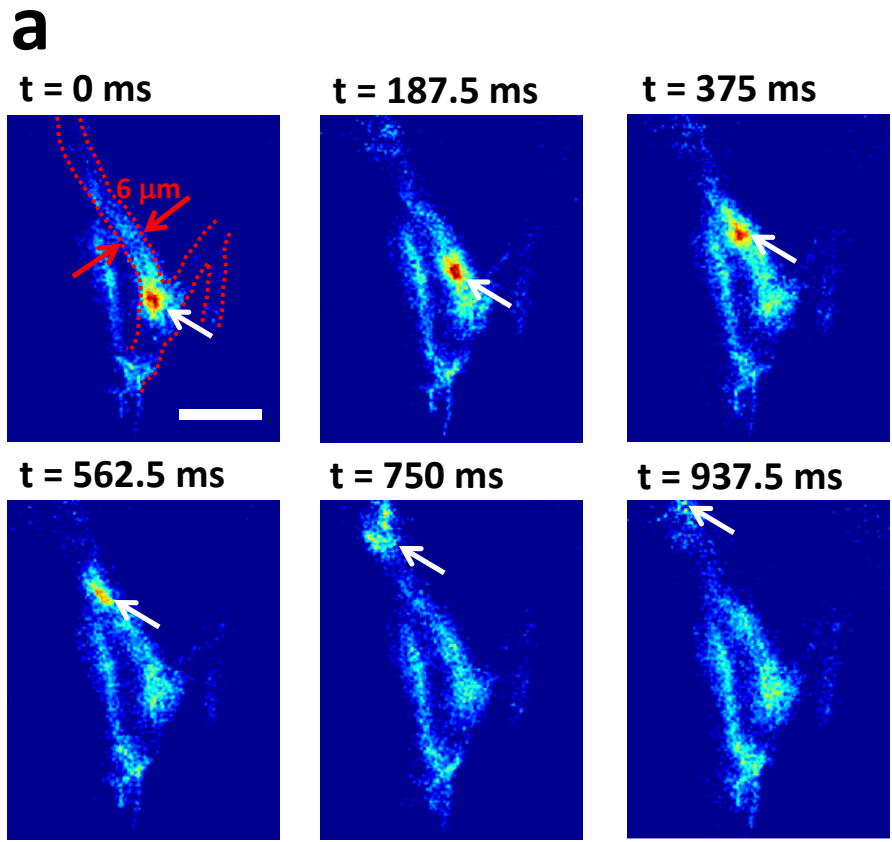

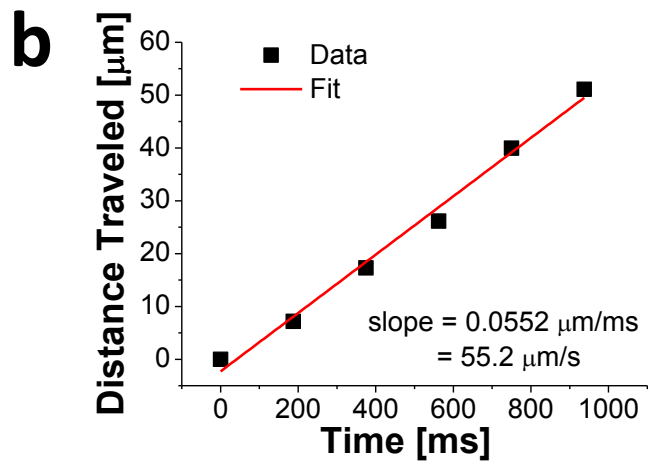

**Supplementary Figure 11 | Blood Flow Tracking Inside a Capillary Vessel.** (**a**) Time course NIR-II fluorescence images of a capillary vessel after injection of pDA-PEG nanoparticles under an excitation of 808 nm. A bolus of injected pDA-PEG travelling upwards along a 6-μm wide capillary vessel (outlined in red dashed lines) can be visualized in the images, as indicated by the white arrows. The scale bar indicates 20 μm. (**b**) A plot of the distance travelled by the bolus of injected pDA-PEG as a function of time. The linear fit reveals an average blood velocity of 55.2 μm/s in this particular capillary vessel, ~800× slower than in the femoral artery and also much less affected by the cardiac output cycles.

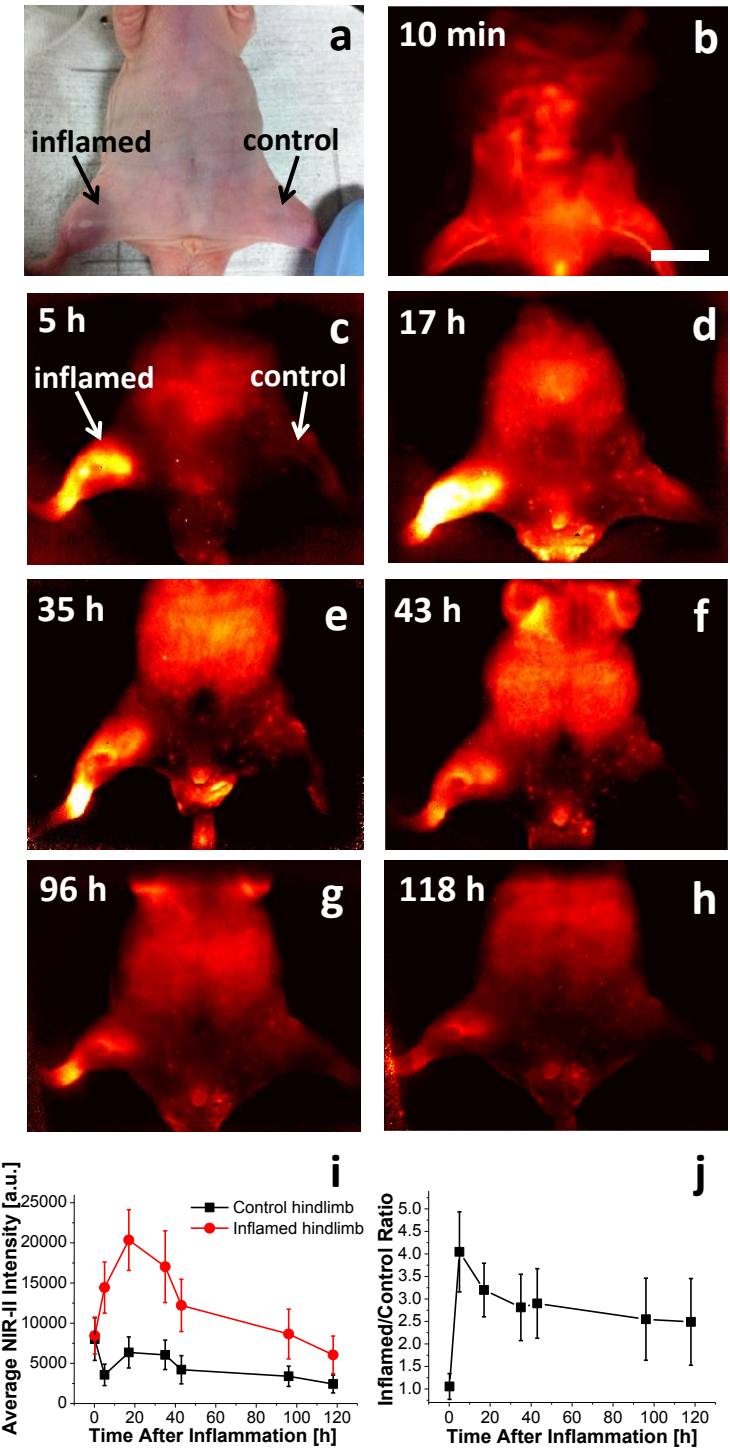

**Supplementary Figure 12 | Imaging Regional Blood Redistribution with pDA-PEG.** (**a**) A white light image showing the mouse with heat-induced inflammation in the right hindlimb (left to the viewer). This image was taken at ~5 h after the inflammation was induced. (**b**-**h**) Time course NIR-II fluorescence images of the mouse at different times after the inflammation was induced. The scale bar in **b** indicates 10 mm and applies to all NIR-II fluorescence images. (**i**) Average NIR-II fluorescence intensity plotted as a function of time after induced inflammation in the control and the inflamed hindlimbs. The error bars in **i** were obtained by taking the

standard deviation of the NIR-II fluorescence intensity of all pixels in each hindlimb at every time point after induced inflammation. (**j**) The ratio of the NIR-II fluorescence intensity in the inflamed hindlimb over that in the control hindlimb, plotted as a function of time after the induced inflammation. The error bars in **j** were obtained by propagating the errors associated with the corresponding control and inflamed hindlimbs shown in **i**.

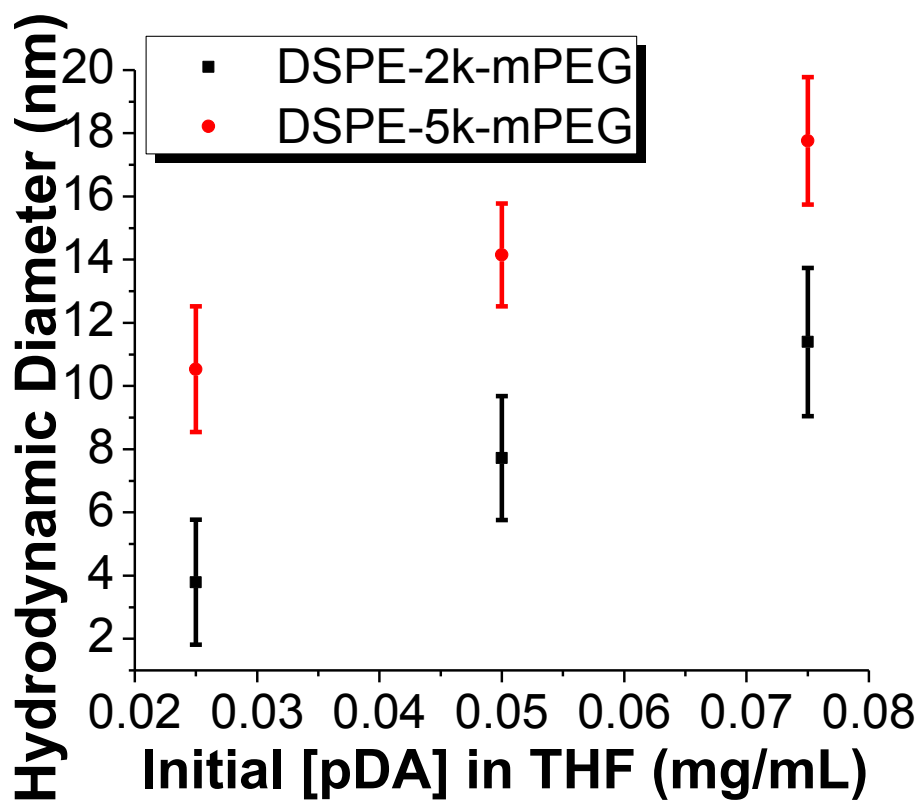

**Supplementary Figure 13 | Size Tunability of the pDA-PEG Nanoparticles.** This plot shows the size distribution of the pDA-PEG nanoparticles measured by DLS and plotted as a function of initial pDA concentration in the THF solution during the synthesis of pDA-PEG nanoparticles and the molecular weight of the surfactant DSPE-mPEG. Note that the measured hydrodynamic diameters of pDA-2k-PEG in the range of 2-6 nm (the bottom left data point) may contain some 'empty' surfactant micelles without any pDA molecules loaded inside (2-4 nm in diameter). The error bars reflect the standard deviation of the measured hydrodynamic diameter distribution of each corresponding sample.

**Supplementary Note 1. | The dynamic range of blood velocity measurement with the NIR-II fluorescence of pDA-PEG.** The dynamic range of velocity measurement using the pDA-PEG fluorophores is from 0 to ~640 mm/s, determined by both the optical properties (the NIR-II fluorescence brightness) of pDA-PEG and the speed of data acquisition of the camera used for dynamic imaging. The detailed derivation of the dynamic range is shown as follows.

Since velocity is defined as distance / time, we need to find out the dynamic ranges of available time intervals of image acquisition and the measurable distance in our imaging system. The frame rate of dynamic fluorescence imaging (i.e., the inverse of time interval per image acquisition), which determines how fast we can track the blood flow *in vivo*, is given by 1 / (exposure time + overhead time). Here the exposure time needed for acquiring images with good quality is determined by the brightness of the pDA fluorophores, and the overhead time is determined by the rate at which the camera acquires and digitizes each image. The shortest possible exposure time is 20 ms (using the brightest pDA polymer we currently have), while the shortest possible overhead time of the camera is ~19 ms. Therefore, the maximum frame rate we can use for dynamic imaging is 1 / (20 ms + 19 ms) = 25.6 Hz (or 25.6 fps). Our imaging system is a low-pass system, meaning any frequency lower than 25.6 Hz is available by increasing the time interval between measurements. Therefore the dynamic range of temporal resolution is from 0 to 25.6 Hz.

On the other hand, by using different lens/objective sets to form images with different magnifications onto the camera, the measurable distance during dynamic imaging ranges from 1 µm (diffraction limit) to 25 mm (field of view limit). Since velocity is distance divided by time, one can derive the slowest and the fastest measurable blood velocity in our current setup using the pDA fluorophores as follows:

The slowest measurable blood velocity = 1 µm $\times$ 0 Hz = 0 mm/s

The fastest measurable blood velocity = 25 mm $\times$ 25.6 Hz = 640 mm/s

Therefore the dynamic range of measurable blood velocity in our current system using the pDA fluorophores is given as 0 to 640 mm/s.